\documentclass[10pt,journal,compsoc]{IEEEtran}
\usepackage[linesnumbered, ruled, vlined]{algorithm2e}
\usepackage{graphicx}
\usepackage{balance}
\usepackage{caption}
\usepackage{moresize}
\usepackage{pbox}
\usepackage{mathtools}
\usepackage[TABBOTCAP]{subfigure}
\usepackage{paralist}
\usepackage{tabularx}
\usepackage{hyperref}
\usepackage{balance}
\usepackage{multirow}
\usepackage{multicol}
\usepackage{amsmath}
\usepackage[usenames,dvipsnames,table]{xcolor}
\usepackage{listings}
\usepackage{setspace}
\usepackage{verbatim}
\usepackage{float}
\usepackage[normalem]{ulem}
\usepackage{xspace}
\usepackage{amssymb}
\usepackage{ifthen}
\usepackage{pifont}
\usepackage{pict2e,picture}

\newcommand{\ReDraw}{{\sc \small ReDraw}\xspace}
\newcommand{\Remaui}{{\sc \small Remaui}\xspace}

\newcommand{\CrashScope}{{\sc \small CrashScope}\xspace}
\newcommand{\MonkeyLab}{{\sc \small MonkeyLab}\xspace}
\newcommand{\pixcode}{{\small pix2code}\xspace}

\newcommand{\ReDrawAbs}{{R}{\ssmall E}{D}{\ssmall RAW}\xspace}

\newcommand{\ReDraws}{{\sc ReDraw's}\xspace}

\include{macro}

\ifCLASSOPTIONcompsoc
  \usepackage[nocompress]{cite}
\else
  \usepackage{cite}
\fi
\ifCLASSINFOpdf
\else
\fi
\begin{document}
%
\title{Machine Learning-Based Prototyping of Graphical User Interfaces for Mobile Apps}
%
%
%
%

\author{Kevin~Moran,~\IEEEmembership{Member,~IEEE,}
        Carlos Bernal-C\'ardenas,~\IEEEmembership{Student Member,~IEEE,}\\
		Michael Curcio,~\IEEEmembership{Student Member,~IEEE,}
		Richard Bonett,~\IEEEmembership{Student Member,~IEEE,}
        and~Denys~Poshyvanyk,~\IEEEmembership{Member,~IEEE}
\IEEEcompsocitemizethanks{\IEEEcompsocthanksitem All authors are with the Department
of Computer Science, College of William \& Mary, Williamsburg,
VA, 23185.\protect\\
E-mail: \{kpmoran, cebernal, mjcurcio, rfbonett, denys\}@cs.wm.edu}
\thanks{Manuscript received May 2018;}}

%
%

\markboth{IEEE Transactions on Software Engineering,~Vol.~\#, No.~\#,~2018}%
{Moran \MakeLowercase{\textit{et al.}}: Machine Learning-Based Prototyping of Graphical User Interfaces for Mobile Apps}
%



\IEEEtitleabstractindextext{%
\begin{abstract}
It is common practice for developers of user-facing software to transform a mock-up of a graphical user interface (GUI) 
into code. This process takes place both at an application's inception and in an evolutionary context as GUI changes keep pace with evolving features. Unfortunately, this practice is challenging and time-consuming.  In this paper, we present an approach that automates this process by enabling accurate prototyping of GUIs via three tasks: \textit{detection}, \textit{classification}, and \textit{assembly}. First, logical components of a GUI are \textit{detected} from a mock-up artifact using either computer vision techniques or mock-up metadata. Then, software repository mining, automated dynamic analysis, and deep convolutional neural networks are utilized to accurately \textit{classify} GUI-components into domain-specific types (\eg toggle-button). Finally, a data-driven, K-nearest-neighbors algorithm generates a suitable hierarchical GUI structure from which a prototype application can be automatically \textit{assembled}. We implemented this approach for Android in a system called \ReDrawAbs. Our evaluation illustrates that \ReDrawAbs achieves an average GUI-component classification accuracy of 91\% and assembles prototype applications that closely mirror target mock-ups in terms of visual affinity while exhibiting reasonable code structure. Interviews with industrial practitioners illustrate ReDraw's potential to improve real development workflows.
\end{abstract}

}

\maketitle

\IEEEdisplaynontitleabstractindextext

%
\IEEEpeerreviewmaketitle


\IEEEraisesectionheading{\section{Introduction}\label{sec:introduction}}
\label{sec:intro}

\IEEEPARstart{M}{ost} modern user-facing software applications are GUI-centric, and rely on attractive user interfaces (UI) and intuitive user experiences (UX) to attract customers, facilitate the effective completion of computing tasks, and engage users.  Software with cumbersome or aesthetically displeasing UIs are far less likely to succeed, particularly as companies look to differentiate their applications from competitors with similar functionality.  This phenomena can be readily observed in mobile application marketplaces such as the App Store \cite{apple-app-store}, or Google Play\cite{google-play}, where many competing applications (also known as \textit{apps}) offering similar functionality (\eg task managers, weather apps) largely distinguish themselves via UI/UX \cite{design-importance}.   Thus, an important step in developing any GUI-based application is drafting and prototyping design mock-ups, which facilitates the instantiation and experimentation of UIs in order to evaluate or prove-out abstract design concepts.  In industrial settings with larger teams, this process is typically carried out by dedicated designers who hold domain specific expertise in crafting attractive, intuitive GUIs using image-editing software such as Photoshop \cite{photoshop} or Sketch \cite{sketch}.  These teams are often responsible for expressing a coherent design language across the many facets of a company's digital presence, including websites, software applications and digital marketing materials.  Some components of this design process also tend to carry over to smaller independent development teams who practice design or prototyping processes by creating wireframes or mock-ups to judge design ideas before committing to spending development resources implementing them. After these initial design drafts are created it is critical that they are faithfully translated into code in order for the end-user to experience the design and user interface in its intended form. 

This process (which often involves multiple iterations) has been shown by past work and empirical studies to be challenging, time-consuming, and error prone \cite{Tucker:CSH04,Myers:CHD94,Nguyen:ASE15,Lelli:ICST15, Moran:ICSE18} particularly if the design and implementation are carried out by different teams (which is often the case in industrial settings \cite{Moran:ICSE18}). Additionally, UI/UX teams often practice an iterative design process, where feedback is collected regarding the effectiveness of GUIs at early stages.  Using prototypes would be preferred, as more detailed feedback could be collected; however, with current practices and tools this is typically too costly \cite{Landay:CHI95,Myers:VLHCC08}. 
Furthermore, past work on detecting GUI design violations in mobile apps highlights the importance of this problem from an industrial viewpoint \cite{Moran:ICSE18}. According to a study conducted with Huawei, a major telecommunications company, 71 unique application screens containing 82 design violations resulting from the company's iterative design and development process were empirically categorized using a grounded-theory approach.  This resulted in a taxonomy of mobile design violations spanning three major categories and 14 subcategories and illustrates the difficulties developers can have faithfully implementing GUIs for mobile apps as well as the burden that design violations introduced by developers can place on the overarching development process.  

Many fast-moving startups and fledgling companies attempting to create software prototypes in order to demonstrate ideas and secure investor support would also greatly benefit from rapid application prototyping.  Rather than spending scarce time and resources on iteratively designing and coding user interfaces, an accurate automated approach would likely be preferred.  This would allow smaller companies to put more focus on features and value and less on translating designs into workable application code.  Given the frustrations that front-end developers and designers face with constructing accurate GUIs, there is a clear need for automated support. 

	To help mitigate the difficulty of this process, some modern IDEs, such as XCode \cite{xcode}, Visual Studio \cite{visual-studio}, and Android Studio \cite{android-studio}, offer built-in GUI editors.  However, recent research suggests that using these editors to create complex, high-fidelity GUIs is cumbersome and difficult \cite{Landay:CHI95},
as users are prone to introducing bugs and presentation failures even for simple tasks \cite{Zeidler:INTERACT13}.  Other commercial solutions include offerings for collaborative GUI-design and for interactive previewing of designs on target devices or browsers (displayed using a custom framework, with limited functionality) \cite{mockup-io,proto-io,fluid-ui,marvelapp,pixate,xiffe,mockingbot,flinto,justinmind,protoapp,irise,appypie,supernova-studio}, \revision{but none offer an end-to-end solution capable of automatically translating a mock-up into accurate native code (with proper component types)} for a target platform. It is clear that a tool capable of even partially automating this process could significantly reduce the burden on the design and development processes.

	Unfortunately, automating the prototyping process for GUIs is a difficult task. At the core of this difficulty is the need to bridge a broad abstraction gap that necessitates reasoning accurate user interface code from either pixel-based, graphical representations of GUIs or digital design sketches.  Typically, this abstraction gap is bridged by a developer's domain knowledge.  For example, a developer is capable of recognizing discrete objects in a mock-up that should be instantiated as components on the screen, categorizing them into proper categories based on their intended functionalities, and arranging them in a suitable hierarchical structure such that they display properly on a range of screen sizes. However, even for a skilled developer, this process can be time-consuming and prone to errors \cite{Moran:ICSE18}. Thus, it follows that an approach which automates the GUI prototyping process must bridge this image-to-code abstraction gap.  This, in turn, requires the creation of a model capable of representing the domain knowledge typically held by a developer, and applying this knowledge to create accurate prototypes.

Given that, within a single software domain, the design and functionality of GUIs can vary dramatically, it is unlikely that manually encoded information or heuristics would be capable of fully supporting such complex tasks.  Furthermore, creating, updating, and maintaining such heuristics manually is a daunting task.  Thus, we propose to learn this domain knowledge using a data-driven approach that leverages machine learning (ML) techniques and the GUI information already present in existing apps (specifically screenshots and GUI metadata) acquired via mining software repositories (MSR).

	More specifically, we present an approach that deconstructs the prototyping process into the tasks of: \textit{detection}, \textit{classification}, and \textit{assembly}. The first task involves \textit{detecting} the bounding boxes of atomic elements (\eg GUI-components which cannot be further decomposed) of a user interface from a mock-up design artifact, such as pixel-based images.  This challenge can be solved either by parsing information regarding objects representing GUI-components directly from mock-up artifacts (\eg parsing exported metadata from Photoshop), or using CV techniques to infer objects \cite{Nguyen:ASE15}.  Once the GUI-components from a design artifact have been identified, they need to be \textit{classified} into their proper domain-specific types (\eg button, dropdown menu, progress bar).  This is, in essence, an image classification task, and research on this topic has shown tremendous progress in recent years, mainly due to advancements in deep convolutional neural networks (CNNs) \cite{Krizhevsky:NIPS12,Zeiler:ECCV14,Simonyan:ICLR14,Szegedy:CVPR15,He:CVPR16}.  However, because CNNs are a supervised learning technique, they typically require a large amount of training data, such as the ILSVRC dataset \cite{Russakovsky:JCV15}, to be effective. We assert that automated dynamic analysis of applications mined from software repositories can be applied to collect screenshots and GUI metadata that can be used to \textit{automatically} derive labeled training data. Using this data, a CNN can be effectively trained to classify images of GUI-Components from a mock-up (extracted using the detected bounding boxes) into their domain specific GUI-component types. However, classified images of components are not enough to \textit{assemble} effective GUI code.  GUIs are typically represented in code as hierarchal trees, where logical groups of components are bundled together in containers. We illustrate that an iterative K-nearest-neighbors (KNN) algorithm and CV techniques operating on mined GUI metadata and screenshots can construct realistic GUI-hierarchies that can be translated into code.

	We have implemented the approach described above in a system called \ReDraw for the Android platform. We mined 8,878 of the top-rated apps from Google Play and executed these apps using a fully automated input generation approach (\eg GUI-ripping) derived from our prior work on mobile testing \cite{Moran:ICST16,Linares:MSR15}. During the automated app exploration the GUI-hierarchies for the most popular screens from each app were extracted.  We then trained a CNN on the most popular native Android GUI-component types as observed in the mined screens.  \ReDraw uses this classifier in combination with an iterative KNN algorithm and additional CV techniques to translate different types of mock-up artifacts into prototype Android apps.  We performed a comprehensive set of three studies evaluating \ReDraw aimed at measuring (i) the accuracy of the CNN-based classifier (measured against a baseline feature descriptor and Support Vector Machine based technique), (ii) the similarity of generated apps to mock-up artifacts (both visually and structurally), and (iii) the potential industrial applicability of our system, through semi-structured interviews with mobile designers and developers at Google, Huawei and Facebook. Our results show that our CNN-based GUI-component classifier achieves a top-1 average precision of $91\%$ (\ie when the top class predicted by the CNN is correct), our generated applications share high visual similarity to their mock-up artifacts, the code structure for generated apps is similar to that of real applications, and \ReDraw has the potential to improve and facilitate the prototyping and development of mobile apps with some practical extensions. Our evaluation also illustrates how \ReDraw outperforms other related approaches for mobile application prototyping, \Remaui \cite{Nguyen:ASE15} and pix2code \cite{Beltramelli:arXiv17}.  Finally, we provide a detailed discussion of the limitations of our approach and promising avenues for future research that build upon the core ideas presented.

In summary, our paper makes the following noteworthy contributions:

\begin{itemize}
 \item The introduction of a novel approach for prototyping software GUIs rooted in a combination of techniques drawn from program analysis, MSR, ML, and CV; and an implementation of this approach in a tool called \ReDraw for the Android platform;
\item A comprehensive empirical evaluation of \ReDraw, measuring several complementary quality metrics, offering comparison to related work, and describing feedback from industry professionals regarding its utility;
\item  An online appendix \cite{appendix} showcasing screenshots of generated apps and study replication information; 
\item \revision{As part of implementing \ReDraw we collected a large dataset of mobile application GUI data containing screenshots and GUI related metadata for over 14k screens and over 190k GUI-components;}  
\item Publicly available open source versions of the \ReDraw code, datasets, and trained ML models\cite{appendix}.
\end{itemize}

\vspace{-1.5em}
\section{Background \& Related Work}
\label{sec:related-work}

	In this section we introduce concepts related to the mock-up driven development process referenced throughout the paper, introduce concepts related to deep convolutional neural networks, and survey related work, distilling the novelty of our approach in context.

\vspace{-1em}
\subsection{Background \& Problem Statement}
\label{sec:problem-statement}

	The first concept of a mock-up driven development practice we reference in this paper is that of \textit{mock-up artifacts}, which we define as: 

\vspace{1em}
\noindent \textbf{\textit{Definition 1} - Mock-Up Artifact: }\textit{An artifact of the software design and development process which stipulates design guidelines for GUIs and its content.} 
\vspace{1em}

\noindent In industrial mobile app development, mock-up artifacts typically come in the form of high fidelity images (with or without meta-data) created by designers using software such as Photoshop \cite{photoshop} or Sketch \cite{sketch}. In this scenario, depending on design and development workflows, metadata containing information about the constituent parts of the mock-up images can be exported and parsed from these artifacts \footnote{For example, by exporting Scalable Vector Graphics (\texttt{\footnotesize .svg}) or \texttt{\footnotesize html} formats from Photoshop.}. 
Independent developers may also use screenshots of existing apps to prototype their own apps.  In this scenario, in addition to screenshots of running applications, runtime GUI-information (such as the html DOM-tree of a web app or the GUI-hierarchy of a mobile app) can be extracted to further aid in the prototyping process. However, this is typically \textit{not} possible in the context of mock-up driven development (which our approach aims to support), as executable apps do not exist. 

	The second concept we define is that of \textit{GUI-components} (also commonly called \textit{GUI-widgets}). In this paper, we use the terms \textit{GUI-component} and \textit{component} interchangeably.  We define these as:

\vspace{1em}
\noindent \textbf{\textit{Definition 2} - GUI-Component: }\textit{Atomic graphical elements with pre-defined functionality, displayed within a GUI of a software application.}
\vspace{1em}

\noindent GUI-components have one of several domain dependent types, with each distinct type serving a different functional or aesthetic purpose. For example, in web apps common component types include dropdown menus and checkboxes, just to name a few.  

	The notion of \textit{atomicity} is important in this definition, as it differentiates GUI-components from \textit{containers}.  The third concept we define is that of a \textit{GUI-container}:

\vspace{1em}
\noindent \textbf{\textit{Definition 3} - GUI-Container: }\textit{A logical construct that groups member GUI-components and typically defines spatial display properties of its members.} 
\vspace{1em}

\noindent In modern GUI-centric apps, GUI-components are rarely rendered on the screen using pre-defined coordinates.  Instead, logical groupings of containers form hierarchical structures (or \textit{GUI-hierarchies}).  These hierarchies typically define spatial information about their constituent components, and in many cases react to changes in the size of the display area (\ie \textit{reactive design})\cite{android-ui-development}. For instance, a GUI-component that displays text may span the text according to the dimensions of its container.

	Given these definitions, the problem that we aim to solve in this paper is the following:

\vspace{1em}
\noindent \textbf{Problem Statement: }\textit{Given a mock-up artifact, generate a prototype application that closely resembles the mock-up GUI both visually, and in terms of expected structure of the GUI-hierarchy.} 
\vspace{0.5em}

\noindent As we describe in Sec. \ref{sec:approach}, this problem can be broken down into three distinct tasks including the \textit{detection} and \textit{classification} of GUI-components, and the \textit{assembly} of a realistic GUI-hierarchy and related code. In the scope of this paper, we focus on automatically generating GUIs for mobile apps that are visually and structurally similar (in terms of their GUI hierarchy).  \revision{To accomplish this we investigate the ability of our proposed approach to automatically prototype applications from two types of mock-up artifacts, (i) images of existing applications, and (ii) Sketch~\cite{sketch} mock-ups reverse engineered from existing popular applications. We utilize these types of artifacts as real mockups are typically not available for open source mobile apps and thus could not be utilized in our study. It should be noted that the two types of mock-up artifacts used in this paper may not capture certain ambiguities that exist in mock-ups created during the course of a real software design process. We discuss the implications of this in Sec. \ref{sec:limitations-threats}.} 

\subsubsection{Convolutional Neural Network (CNN) Background}

	In order to help classify images of GUI-components into thier domain specific types, \ReDraw utilizes a Convolutional Neural Network (CNN). To provide background for the unfamiliar reader, in this sub-section we give an overview of a typical CNN architecture, explaining elements of the architecture that enable accurate image classification. However, for more comprehensive descriptions of CNNs, we refer readers to \cite{Krizhevsky:NIPS12} \& \cite{conv-nets}. 

\begin{figure}[t]
	\centering
	\includegraphics[width=1.00\linewidth]{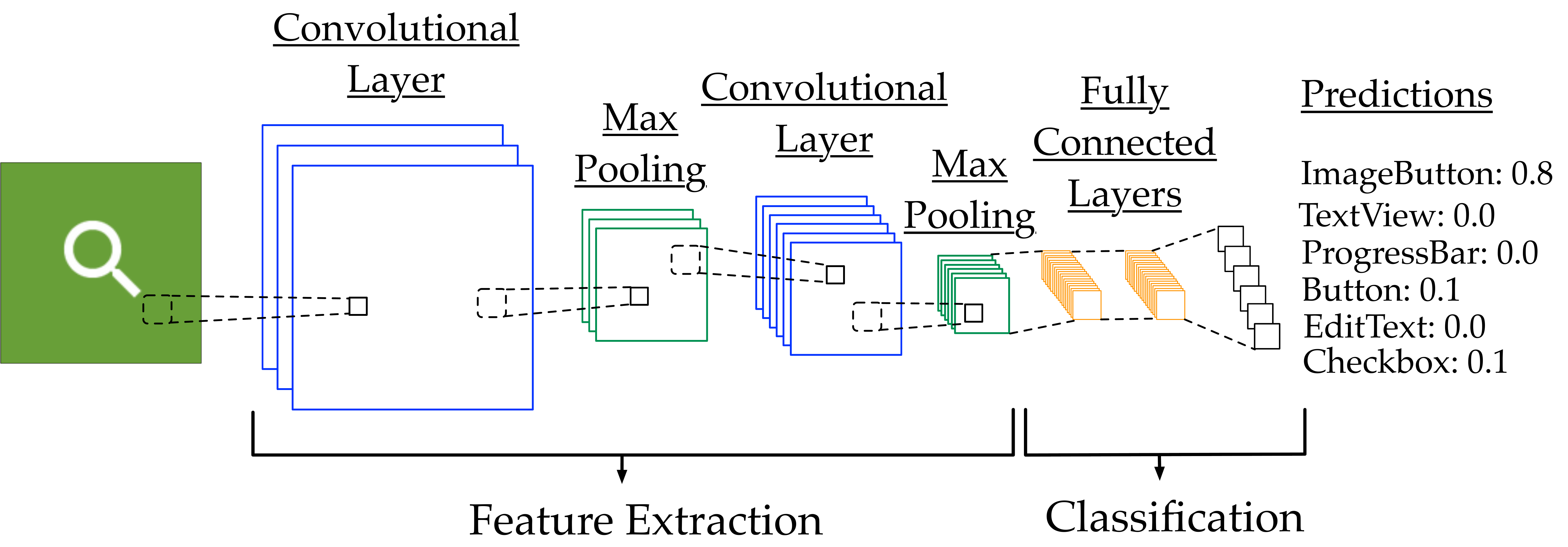}
\vspace{-1.5em}
	\caption{Typical Components of CNN Architecture}
	\label{fig:cnn-architecture}
\vspace{-0.3em}
\end{figure}

	\noindent{\textbf{CNN Overview:}} Fig. \ref{fig:cnn-architecture} illustrates the basic components of a traditional CNN architecture. As with most types of artificial neural networks, CNNs typically encompass several different \textit{layers} starting with an input layer where an image is passed into the network, then to hidden layers where abstract features, and weights representing the ``importance" of features for a target task are learned. CNNs derive their name from unique ``convolutional" layers which operate upon the mathematical principle of a convolution \cite{conv-operator}.  The purpose of the convolutional layers, shown in blue in Figure \ref{fig:cnn-architecture}, are to extract features from images.  Most images are stored as a three (or four) dimensional matrix of numbers, where each dimension of the matrix represents the intensity of a color channel (\eg RGB). Convolutional layers operate upon these matrices using a \textit{filter} (also called kernel, or feature detector), which can be thought of as a sliding window of size $n$ by $m$ that slides across an set of matricies representing an image. This window applies a convolution operation (\ie an element-wise matrix multiplication) creating a \textit{feature map}, which represents extracted image features. As convolution layers are applied in succession, more abstract features are learned from the original image. \textit{Max Pooling} layers also operate as a sliding window, pooling maximum values in the feature maps to reduce dimensionality.  Finally, fully-connected layers and a softmax classifier act as a multi-layer perceptron to perform classification. CNN training is typically performed using gradient descent, and back-propagation of error gradients.

	\noindent{\textbf{Convolutional Layers:}} Convolutional layers extract feature maps from images to learn high level features. The size of this feature map results from three parameters: (i) the number of filters used, (ii) the stride of the sliding window, and (iii) whether or not padding is applied. Leveraging multiple filters allows for multi-dimensional feature maps, the stride corresponds to the distance the sliding window moves during each iteration, and padding can be applied to learn features from the borders of an input image. These feature maps are intended to represent abstract features from images, which inform the prediction process.   

\noindent{\textbf{Rectified Linear Units (ReLUs):}} Traditionally, an element of non-linearity is introduced after each convolutional layer, as the convolution operator is linear in nature, which may not correspond to non-linear nature of data being learned. The typical manner in which this non-linearity is introduced is through \textit{Rectified Linear Units} (ReLUs). The operation these units perform is simple in nature, replacing all negative values in a feature map with zeros. After the convolutions and ReLU operations have been performed, the resulting feature map is typically subjected to \textit{max pooling} (Fig. \ref{fig:cnn-architecture}). 

\noindent{\textbf{Max Pooling:}} Max pooling again operates as as sliding window, but instead of performing a convolution, simply pools the maximum value from each step of the sliding window. This allows for a reduction in the dimensionality of the data while extracting salient features. 

\noindent{\textbf{Fully Connected Layers:}} The layers described thus far in the network have been focused on deriving features from images.  Therefore, the final layers of the network must utilize these features to compute predictions about classes for classifications. This is accomplished via the \textit{fully connected layers}, which act as a multi-layer perceptron typically utilizing a softmax activation function. 

\noindent{\textbf{CNN Training Procedure:}} Training a CNN is accomplished through back-propagation. After the initialization of all the network parameters, initial weights are set to random values.  Then input images are fed through the network layers in the forward direction, and the total error across all output classes is calculated.  This error is back-propagated through the network and \textit{gradient descent} is used to calculate error gradients for the network weights which are then updated to minimize the output error.  A \textit{learning rate} controls the degree to which weights are updated based on the gradient calculations.  This process is repeated over the entire training image set, which allows for training both feature extraction and classification in one automated process. After training is complete, the network should be capable of effective classification of input images.  

\subsection{Related Work}
\label{sec:problem-statement} 

\subsubsection{Reverse Engineering Mobile User Interfaces:} 
\label{subsubsec:remaui}

	The most closely related research to the approach proposed in this paper is \Remaui, which aims to reverse engineer mobile app GUIs \cite{Nguyen:ASE15}. \Remaui uses a combination of Optical Character Recognition (OCR), CV, and mobile specific heuristics to detect components and generate a static app. The CV techniques utilized in \Remaui are powerful, and we build upon these innovations. However, \Remaui has key limitations compared to our work including: (i) it does not support the classification of detected components into their native component types and instead uses a binary classification of either text or images, limiting the real-world applicability of the approach, and (ii) it is unclear if the GUI-hierarchies generated by \Remaui are realistic or useful from a developer's point of view, as the GUI-hierarchies of the approach were not evaluated.  
	
	In comparison, the approach presented in this paper (i) is not specific to any particular domain (although we implement our approach for the Android platform as well) as we take a data-driven approach for classifying and generating GUI-hierarchies, (ii) is capable of classifying GUI-components into their respective types using a CNN, and (iii) is able to produce realistic GUI-hierarchies using a data-driven, iterative KNN algorithm in combination with CV techniques. In our evaluation, we offer a comparison of \ReDraw to the \Remaui approach according to different quality attributes in Sections \ref{sec:study} \& \ref{sec:results}.
 
In addition to \Remaui, an open access paper (\ie non-peer-reviewed) was recently posted that implements an approach called pix2code \cite{Beltramelli:arXiv17}, which shares common goals with the research we present in this paper.  Namely, the authors implement an encoder/decoder model that they trained on information from GUI-metadata and screenshots to translate target screenshots first into a domain specific language (DSL) and then into GUI code.  However, this approach exhibits several shortcomings that call into question the real-world applicability of the approach: (i) the approach was only validated on a small set of synthetically generated applications, and no large-scale user interface mining was performed; (ii) the approach requires a DSL which will need to be maintained and updated over time, adding to the complexity and effort required to utilize the approach in practice. Thus, it is difficult to judge how well the approach would perform on real GUI data.  In contrast, \ReDraw is trained on a large scale dataset collected through a novel application of automated dynamic analysis for user interface mining. The data-collection and training process can be performed completely automatically and iteratively over time, helping to ease the burden of use for developers. To make for a complete comparison to current research-oriented approaches, we also include a comparison of the prototyping capability for real applications between \ReDraw and the pix2code approach in Sections \ref{sec:study} \& \ref{sec:results}. 

\revision{
\vspace{-0.3cm}
\subsubsection{Mobile GUI Datasets}
\label{subsubsec:mobile-gui-datasets} 

	In order to train an accurate CNN classifier, \ReDraw requires a large number of GUI-component images labeled with their domain specific types. In this paper, we collect this dataset in a completely automated fashion by mining and automatically executing the top-250 Android apps in each category of Google Play excluding game categories, resulting in 14,382 unique screens and 191,300 labeled GUI-components (after data-cleaning). Recently, (while this paper was under review) a large dataset of GUI-related information for Android apps, called RICO, was published and made available \cite{Deka:UIST17}. This dataset is larger than the one collected in this paper, containing over 72k unique screens and over 3M GUI-components. However, the \ReDraw dataset is differentiated by some key factors specific to the problem domain of prototyping mobile GUIs:

\begin{enumerate}

	\item{\textbf{\textit{Cropped Images of GUI-components:}} The \ReDraw dataset of mobile GUI data contains a set of labeled GUI-components cropped from larger screenshots that are ready for processing by machine learning classifiers.}
	\item{\textbf{\textit{Cleaned Dataset:}} We implemented several filtering procedures at the app, screen, and GUI-component level to remove ``noisy" components from the \ReDraw dataset. This is an important factor for training an effective, accurate machine-learning classifier. These filtering techniques were manually verified for accuracy.}
	\item{\textbf{\textit{Data Augmentation:}} In the extraction of our dataset, we found that certain types of components were used more often than others, posing problems for deriving a \textit{balanced} dataset of GUI-component types. To help mitigate this problem, we utilized data-augmentation techniques to help balance our observed classes.}
\end{enumerate}

	We expand on the methodology for deriving the \ReDraw dataset in Section \ref{subsec:impl-classification}. The RICO dataset does not exhibit the unique characteristics of the \ReDraw dataset stipulated above that cater to creating an effective machine-learning classifier for classifying GUI-components.  However, it should be noted that future work could adapt the data cleaning and augmentation methodologies stipulated in this paper to the RICO dataset to produce a larger training set for GUI-components in the future. 

}

\subsubsection{Other GUI-Design and Reverse Engineering Tools:}	
\label{subsubsec:related-approaches} 

	Given the prevalence of GUI-centric software, there has been a large body of work dedicated to building advanced tools to aid in the construction of GUIs and related code \cite{Coyette:INTERACT07,Caetano:02,Landay:IEEE01,Chatty:UIST04,Seifert:MobileHCI11,Meng:CHI14,Lasecki:CHI15} and to reverse engineer GUIs \cite{Chang:UIST11,Dixon:CHI11,Dixon:CHI10,Hinze:EICS10,Shah:SPLASH11,Samir:ICICT07}.  While these approaches are aimed at various goals, they all attempt to reason logical, or programatic info from graphical representations of GUIs.
	
	However, the research projects referenced above exhibit one or more of the following attributes: \revision{(i) they do not specifically aim to support the task of automatically translating existing design mock-ups into code} \cite{Dixon:CHI11,Dixon:CHI10,Hinze:EICS10,Shah:SPLASH11,Samir:ICICT07}, (ii) they force designers or developers to compromise their workflow by imposing restrictions on how applications are designed or coded \cite{Seifert:MobileHCI11,Meng:CHI14,Landay:IEEE01,Caetano:02,Coyette:INTERACT07,Chatty:UIST04} or (iii) they rely purely on reverse engineering existing apps using runtime information, which is not possible in the context of mock-up driven development~\cite{Chang:UIST11,Meng:CHI14}.
\revision{These attributes indicate that the above approaches are either not applicable in the problem domain described in this paper (automatically generating application code from a mock-up artifact) or represent significant limitations that severely hinder practical applicability.}  Approaches that tie developers or designers into strict workflows (such as restricting the ways mock-ups are created or coded) struggle to gain adoption due to the competing flexibility of established image-editing software and coding platforms. Approaches requiring runtime information of a \textit{target} app cannot be used in a typical mock-up driven development scenario, as implementations do not exist yet. While our approach relies on runtime data, it is collected and processed \textit{independently} of the target app or mock-up artifact. Our approach aims to overcome the shortcomings of previous research by leveraging MSR and ML techniques to automatically infer models of GUIs for different domains, and has the potential to integrate into current design workflows as illustrated in Sec. \ref{subsec:results-rq4}.

\begin{figure*}[t]
  \centering

  \includegraphics[width=\linewidth]{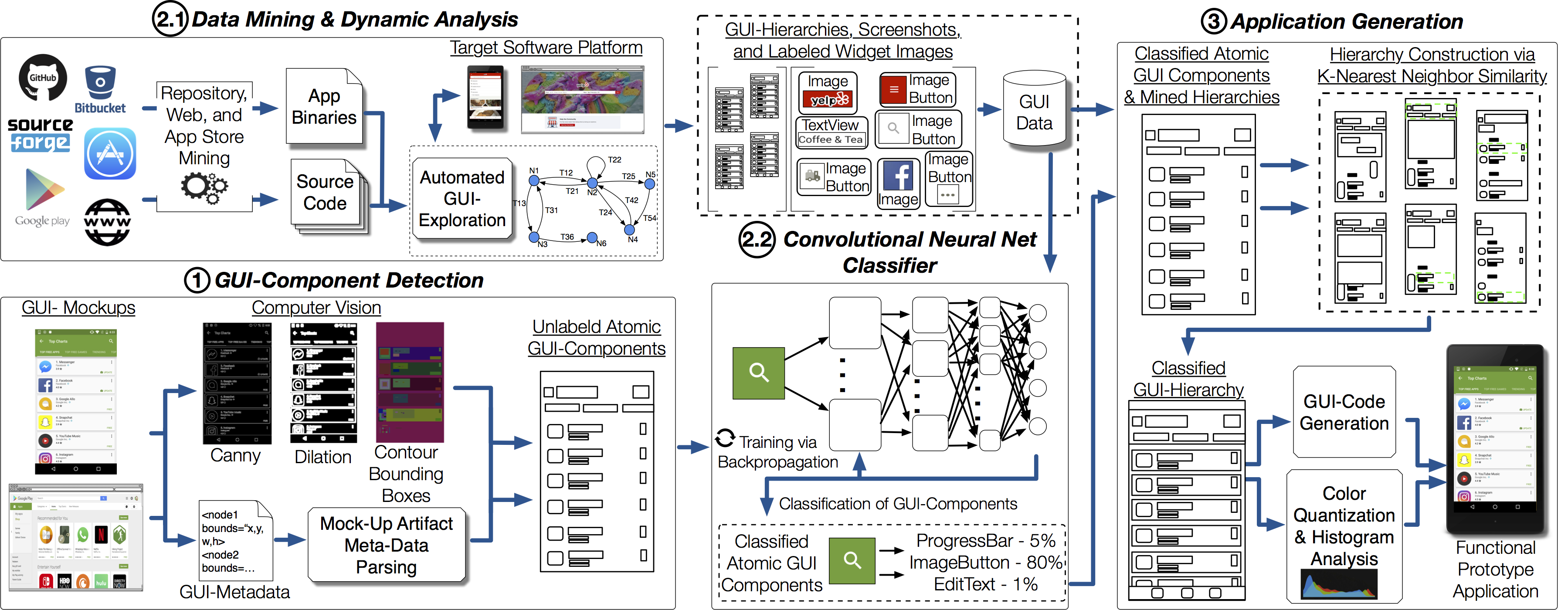}

  \caption{Overview of Proposed Approach for Automated GUI-Prototyping}

  \label{fig:framework-overview}
\end{figure*}

	In addition to research on this topic, there are several commercial solutions which aim to improve the mock-up and prototyping process for different types of applications \cite{mockup-io,proto-io,fluid-ui,marvelapp,pixate,xiffe,mockingbot,flinto,justinmind,protoapp,irise,appypie}.  These approaches allow for better collaboration among designers, and some more advanced offerings enable limited-functionality prototypes to be displayed on a target platform with support of a software framework. For instance, some tools will display screenshots of mock-ups on a mobile device through a preinstalled app, and allow designers to preview designs.  However, these techniques \textit{are not} capable of translating mock-up artifacts into GUI code, and tie designers into a specific, potentially less flexible software or service. \revision{While this paper was under review, a recent startup has released software called Supernova Studio~\cite{supernova-studio} that claims to be able to translate Sketch files into native code for iOS and Android. While this platform does contain some powerful features, such as converting Sketch screen designs into GUI code with ``reactive" component coordinates, it exhibits two major drawbacks: (i) it is inherently tied to the Sketch application, and does not allow imports from other design tools, and (ii) it is not capable of classifying GUI-components into their respective types, instead relying on a user to complete this process manually~\cite{supernova-studio-limitations}. Thus, \ReDraw is complementary in the sense that our GUI-component classification technique could be used in conjunction with Supernova Studio to improve its overall effectiveness.} 

\subsubsection{Image Classification using CNNs:	}
\label{subsubsec:related-approaches} 

	  Large scale image recognition and classification has seen tremendous progress mainly due to advances in CNNs\cite{Krizhevsky:NIPS12,Zeiler:ECCV14,Simonyan:ICLR14,Szegedy:CVPR15,He:CVPR16,Lecun:IEEE98}. These supervised ML approaches are capable of automatically learning robust, salient features of image categories from large numbers of labeled training images such as the ILSVRC dataset \cite{Russakovsky:JCV15}. Building on top of LeCun's pioneering work \cite{Lecun:IEEE98}, the first approach to see a significant performance improvement over existing techniques (that utilized predefined feature extraction) was AlexNet \cite{Krizhevsky:NIPS12}, which achieved a top-5 mean average error (MAE) of $\approx 15\%$ on ILSVRC12. The architecture for this network was relatively shallow, but later work would show the benefits and tradeoffs of using deeper architectures.  Zeiler and Fergus developed the ZFNet \cite{Zeiler:ECCV14} architecture which was able to achieve a lower top-5 MAE than AlexNet ($\approx 11\%$) and devised a methodology for visualizing the hidden layers (or activation maps) of CNNs. More recent approaches such as GoogLeNet \cite{Szegedy:CVPR15} and Microsoft's ResNet \cite{He:CVPR16} use deeper architectures (e.g., 22 and 152 layers respectively) and have managed to surpass human levels of accuracy on image classification tasks.  However, the gains in network learning capacity afforded by deeper architectures come with a trade off in terms of training data requirements and training time.  In this paper, we show that a relatively simple CNN architecture can be trained in a reasonable amount of time on popular classes of Android GUI-components, achieving a top-1 average classification accuracy of $91\%$.


\vspace{-1em}
\section{Approach Description}
\label{sec:approach}

	We describe our approach for GUI prototyping around the three major phases of the process: \textit{detection}, \textit{classification}, \& \textit{assembly}.  Fig. \ref{fig:framework-overview} illustrates an overview of the process that we will refer to throughout the description of the approach.   At a high-level, our approach first \textit{detects} GUI-components from a mock-up artifact by either utilizing CV techniques or parsing meta-data directly from mock-up artifacts generated using professional photo-editing software.  Second, to \textit{classify} the detected GUI-components into proper types, we propose to train a CNN using GUI data gleaned from large-scale automated dynamic analysis of applications extracted by mining software repositories.  The trained CNN can then be applied to mock-up artifacts to classify detected components.  Finally, to construct a suitable GUI-hierarchy (\eg proper groupings of GUI-components in GUI-containers) we utilize a KNN-based algorithm that leverages the GUI-information extracted from the large-scale dynamic analysis to \textit{assemble} a realistic nested hierarchy of GUI-components and GUI-containers. To illustrate our general approach, for each phase we first describe the proposed methodology and design decisions at a high level and then discuss the implementation details specific to our instantiation of \ReDraw for the Android platform.

\subsection{Phase 1 - Detection of GUI-Components}
\label{subsec:research-task-1}

	The first task required of a GUI-prototyping approach is detecting the GUI-components that exist in a mock-up artifact. The main \textit{goal} of this phase is to accurately infer the bounding boxes of atomic GUI-component elements (in terms of pixel-based coordinates) from a mock-up artifact.  This allows individual images of GUI-components to be cropped and extracted in order to be utilized in the later stages of the prototyping process.  This phase can be accomplished via one of two methodologies: (i) parsing data from mock-up artifacts, or (ii) using CV techniques to detect GUI-components.  A visualization of this phase is illustrated in Fig. \ref{fig:framework-overview}-\circled{1}. In the following subsections we describe the \textit{detection} procedure for both of these methodologies as well as our specific implementation within \ReDraw.

\subsubsection{Parsing Data from Design Mockups}
\label{subsec:parsing-mockups}

	The first method for detecting the GUI-components that exist in a mock-up artifact, shown in the bottom portion of Fig. \ref{fig:framework-overview}-\circled{1}, is to utilize the information encoded into mock-up artifacts.  Given the importance of UI/UX in today's consumer facing software, many designers and small teams of developers work with professional grade image editing software, such as Photoshop \cite{photoshop} or Sketch \cite{sketch} to create either wireframe or pixel perfect static images of GUIs that comprise mock-up artifacts. During this process photo-editing or design software is typically used to create a blank canvas with dimensions that match a target device screen or display area (with some design software facilitating scaling to multiple screen sizes \cite{photoshop,sketch}).  Then, images representing GUI-components are placed as editable objects on top of this canvas to construct the mock-up. Most of these tools are capable of exporting the mock-up artifacts in formats that encode spatial information about the objects on the canvas, such as using the Scalable Vector Graphics (\texttt{\small .svg}) format or \texttt{html} output \cite{marketch}. Information about the layouts of objects, including the bounding boxes of these objects, can be parsed from these output formats, resulting in highly accurate detection of components.  Therefore, if this metadata for the mock-up artifacts is available, it can be parsed to obtain extremely accurate bounding boxes for GUI-components that exist in a mock-up artifact which can then be utilized in the remainder of the prototyping process.  

	Given the spatial information encoded in metadata that is sometimes available in mock-up artifacts, one may question whether this information can also be used to reconstruct a hierarchical representation of GUI-components that could later aid in the code conversion process.  Unfortunately, realistic \textit{GUI-hierarchies} typically cannot be feasibly parsed from such artifacts for at least the following two reasons: (i) designers using photo-editing software to create mock-ups tend to encode a different hierarchal structure than a developer would, due to a designer lacking knowledge regarding the best manner in which to programmatically arrange GUI-components on a screen \cite{Moran:ICSE18}; (ii) limitations in photo-editing software can prohibit the creation of programmatically proper spatial layouts.  Thus, any hierarchical structure parsed out of such artifacts is likely to be specific to designers' preferences, or restricted based  on the capabilities of photo-editing software, limiting applicability in our prototyping scenario.  For example, a designer might not provide enough GUI-containers to create an effective reactive mobile layout, or photo-editing software might not allow for relative positioning of GUI-components that scale across different screen sizes.

\subsubsection{Using CV Techniques for GUI-component Detection:}
\label{subsec:cv-detection}

	While parsing information from mock-ups results in highly accurate bounding boxes for GUI-components this info may not always be available, either due to limitations in the photo-editing software being used or differing design practices, such as digitally or physically sketching mockups using pen displays, tablets, or paper.  In these cases, a mock-up artifact may consist only of an image, and thus CV techniques are needed to identify relevant GUI-component info. To support these scenarios, our approach builds upon the CV techniques from \cite{Nguyen:ASE15} to detect GUI-component bounding boxes. This process uses a series of different CV techniques (Fig. \ref{fig:framework-overview}-\circled{1}) to infer bounding boxes around objects corresponding to GUI components in an image.  First, Canny's edge detection algorithm \cite{Canny:TPAMI86} is used to detect the edges of objects in an image.  Then these edges are dilated to merge edges close to one another. Finally, the contours of those edges are used to derive bounding boxes around atomic GUI-components.  Other heuristics for merging text-based components using Optical Character Recognition (OCR) are used to merge the bounding boxes of logical blocks of text (\eg rather than detecting each word as its own component, sentences and paragraphs of text are merged).

\subsubsection{ReDraw Implementation - GUI Component Detection}
\label{subsec:impl-detection}  

	In implementing \ReDraw, to support the scenario where metadata can be gleaned from mock-ups for Android applications we target artifacts created using the Marketch \cite{marketch} plugin for Sketch \cite{sketch}, which exports mock-ups as a combination of \texttt{\small html} \& \texttt{\small javascript}. Sketch is popular among mobile developers and offers extensive customization through a large library of plugins \cite{sketch-ext}. \ReDraw parses the bounding boxes of GUI-components contained within the exported Marketch files. 

	To support the scenario where meta-data related to mock-ups is not available, \ReDraw uses CV techniques to automatically infer the bounding boxes of components from a static image.  To accomplish this, we re-implemented the approach described in \cite{Nguyen:ASE15}. Thus, the input to the GUI-component detection phase of \ReDraw is either a screenshot and corresponding \texttt{marketch} file (to which the marketch parsing procedure is applied), or a single screenshot (to which CV-based techniques are applied). The end result of the GUI-component detection process is a set of bounding box coordinates situated within the original input screenshot and a collection of images cropped from the original screenshot according to the derived bounding boxes that depict atomic GUI-components. This information is later fed into a CNN to be classified into Android specific component types in Phase 2.2. It should be noted that only \textit{GUI-components} are detected during this process. On the other hand \textit{GUI-containers} and the corresponding GUI-hierarchy are constructed in the \textit{assembly} phase described in Sec. \ref{subsec:research-task-3}.

\subsection{Phase 2 - GUI-component Classification}
\label{subsec:research-task-2.1}

	Once the bounding boxes of atomic GUI-component elements have been \textit{detected} from a mock-up artifact, the next step in the prototyping process is to \textit{classify} cropped images of specific GUI components into their domain specific types. To do this, we propose a data-driven and ML-based approach that utilizes CNNs.  As illustrated in Fig. \ref{fig:framework-overview}-{\footnotesize \circledlong{2.1}} and Fig. \ref{fig:framework-overview}-{ \footnotesize \circledlong{2.2}}, this phase has two major parts: (i) large scale software repository mining and automated dynamic analysis, and (ii) the training and application of a CNN to classify images of GUI-components. In the following subsections we first discuss the motivation and implementation of the repository mining and dynamic analysis processes before discussing the rationale for using a CNN and our specific architecture and implementation within \ReDraw.

\subsubsection{Phase 2.1 - Large-Scale Software Repository Mining and Dynamic Analysis}
\label{subsec:research-task-2.1}

	Given their supervised nature and deep architectures, CNNs aimed at the image classification task require a large amount of training data to achieve precise classification.  Training data for traditional CNN image classification networks typically consists of a large set of images labeled with their corresponding classes, where labels correspond to the primary subject in the image.  Traditionally, such datasets have to be manually procured, wherein humans painstakingly label each image in the dataset. However, we propose a methodology that \textit{automates} the creation of labeled training data consisting of images of specific GUI-components cropped from full screenshots and labels corresponding to their domain specific type (\eg Buttons, or Spinners in Android) using fully-automated dynamic program analysis.

	Our key insight for this automated dynamic analysis process is the following: \textit{during automated exploration of software mined from large repositories, platform specific frameworks can be utilized to extract meta-data describing the GUI, which can then be transformed into a large labeled training set suitable for a CNN.}  As illustrated in Fig. \ref{fig:framework-overview}-{\footnotesize \circledlong{2.1}}, this process can be automated by mining software repositories to extract executables.  Then a wealth of research in automated input generation for GUI-based testing of applications can be used to automatically execute mined apps by simulating user-input. For instance, if the target is a mobile app, input generation techniques relying on random-based \cite{Machiry:FSE13,android-monkey,intent-fuzzer,Sasnauskas:WODA14,Ye:MoMM13}, systematic \cite{Azim:OOPSLA13,Anand:FSE12,Amalfitano:ASE12,Moran:ICST16,Moran:ICSE17}, model-based \cite{Amalfitano:ASE12,Yang:FASE13,Azim:OOPSLA13,Choi:OOPSLA13,Hao:Mobisys14,Zaeem:ICST2014,Linares:MSR15}, or evolutionary \cite{Mao:ISSTA16,Mahmood:FSE14} strategies could be adopted for this task.  As the app is executed, screenshots and GUI-related metadata can be automatically extracted for each unique observed screen or layout of an app.  Other similar automated GUI-ripping or crawling approaches can also be adapted for other platforms such as the web \cite{RoyChoudhary:2014:XWA:2610384.2628057,Thome:2014:SST:2593833.2593835,RoyChoudhary:ICSM10,Choudhary:ICST12,NguyenASE2013}. 

	Screenshots can be captured using third party software or utilities included with a target operating system.  GUI-related metadata can be collected from a variety of sources including accessibility services \cite{Grechanik:ICSTW09}, \texttt{\small html} DOM information, or UI-frameworks such as \texttt{\small uiautomator} \cite{uiautomator}. The GUI-metadata and screenshots can then be used to extract sub-images of GUI-components with their labeled types parsed from the related metadata describing each screen.  The underlying quality of the resulting dataset relates to how well the labels describe the type of GUI-components displayed on a screen.  Given that many of the software UI-frameworks that would be utilized to mine such data pull their information directly from utilities that render application GUI-components on the screen, this information is likely to be highly accurate. However, there are certain situations where the information gleaned from these frameworks contains minor inaccuracies or irrelevant cases.  We discuss these cases and steps that can be taken to mitigate them in Sec. \ref{subsec:impl-classification}.

\subsubsection{ReDraw Implementation - Software Repository Mining and Automated Dynamic Analysis}
\label{subsubsec:impl-app-mining}

	To procure a large set of Android apps to construct our training, validation, and test corpora for our CNN we mined free apps from Google Play at scale.  To ensure the representativeness and quality of the apps mined, we extracted all categories from the Google Play store as of June 2017.  Then we filtered out any category that primarily consisted of games, as games tend to use non-standard types of GUI-components that cannot be automatically extracted.  This left us with a total of 39 categories. We then used a Google Play API library \cite{google-api} to download the top 240 \texttt{\small APKs} from each category, excluding duplicates that existed in more than one category.  This resulted in a total of 8,878 unique \texttt{\small APKs} after accounting for duplicates cross-listed across categories. 

	To extract information from the mined \texttt{\small APKs}, we implemented a large-scale dynamic analysis engine, called the \textit{Execution Engine} that utilizes a systematic automated input generation approach based on our prior work on \CrashScope and \MonkeyLab \cite{Linares:MSR15,Moran:FSE15,Moran:ICST16,Moran:ICSE17} to explore the apps and extract screenshots and GUI-related information for visited screens. More specifically, our systematic GUI-exploration navigates a target apps's GUI in a Depth-First-Search (DFS) manner to exercise tappable, long-tappable, and type-able (\eg capable of accepting text input) components. During the systematic exploration we used Android's \texttt{\small uiautomator} framework \cite{uiautomator} to extract GUI-related info as \texttt{\small xml} files that describe the hierarchy and various properties of components displayed on a given screen. We used the Android \texttt{\small screencap} utility to collect screenshots. The \texttt{\small uiautomator xml} files contain various attributes and properties of each GUI-component displayed on an Android application screen, including the bounding boxes (\eg precise location and area within the screen) and component types (\eg EditText, Toggle Button).  These attributes allow for individual sub-images for each GUI-component displayed on a given screen to be extracted from the corresponding screenshot and automatically labeled with their proper type.  

	The implementation of our DFS exploration strategy utilizes a state machine model where states are considered unique app screens, as indicated by their activity name and displayed window (\eg dialog box) extracted using the \texttt{\small adb shell dumpsys window} command. To allow for feasible execution times across the more than $8.8k$ apps in our dataset while still exploring several app screens, we limited our exploration strategy to exercising 50 actions per app. Prior studies have shown that most automated input generation approaches for Android tend to reach near-peak coverage \revision{(\eg between $\approx 20\textnormal{ and } 40\%$ statement coverage)} after ~5 minutes of exploration \cite{Choudhary:ASE15}.  While different input generation approaches tend to exhibit different numbers of actions per given unit of time, our past work shows that our automated input generation approach achieves competitive coverage to similar approaches \cite{Moran:ICST16}, and our stipulation of 50 actions comfortably exceeds 5 minutes per app.  Furthermore, our goal with this large scale analysis was not to completely explore each application, but rather ensure a diverse set of screens and GUI-Component types.  For each app the \textit{Execution Engine} extracted \texttt{\small uiautomator} files and screenshot pairs for the top six unique screens of each app based on the number of times the screen was visited. If fewer than six screens were collected for a given app, then the information for all screens was collected.  Our large scale \textit{Execution Engine} operates in a parallel fashion, where a centralized dispatcher allocated jobs to workers, where each worker is connected to one physical Nexus 7 tablet and is responsible for coordinating the execution of incoming jobs.  During the dynamic analysis process, each job consists of the systematic execution of a single app from our dataset. When a worker finished with a job, it then notified the dispatcher which in turn allocates a new job.  This process proceeded in parallel across 5 workers until all applications in our dataset had been explored.  Since Ads are popular in free apps \cite{Ruiz:IEEE14,Gui:ICSE15}, and are typically made up of dynamic \textit{WebViews} and not native components, we used \textsl{Xposed} \cite{xposed} to block Ads in apps that might otherwise obscure other types of native components.

\begin{figure}[t]
	\centering
	\includegraphics[width=\linewidth]{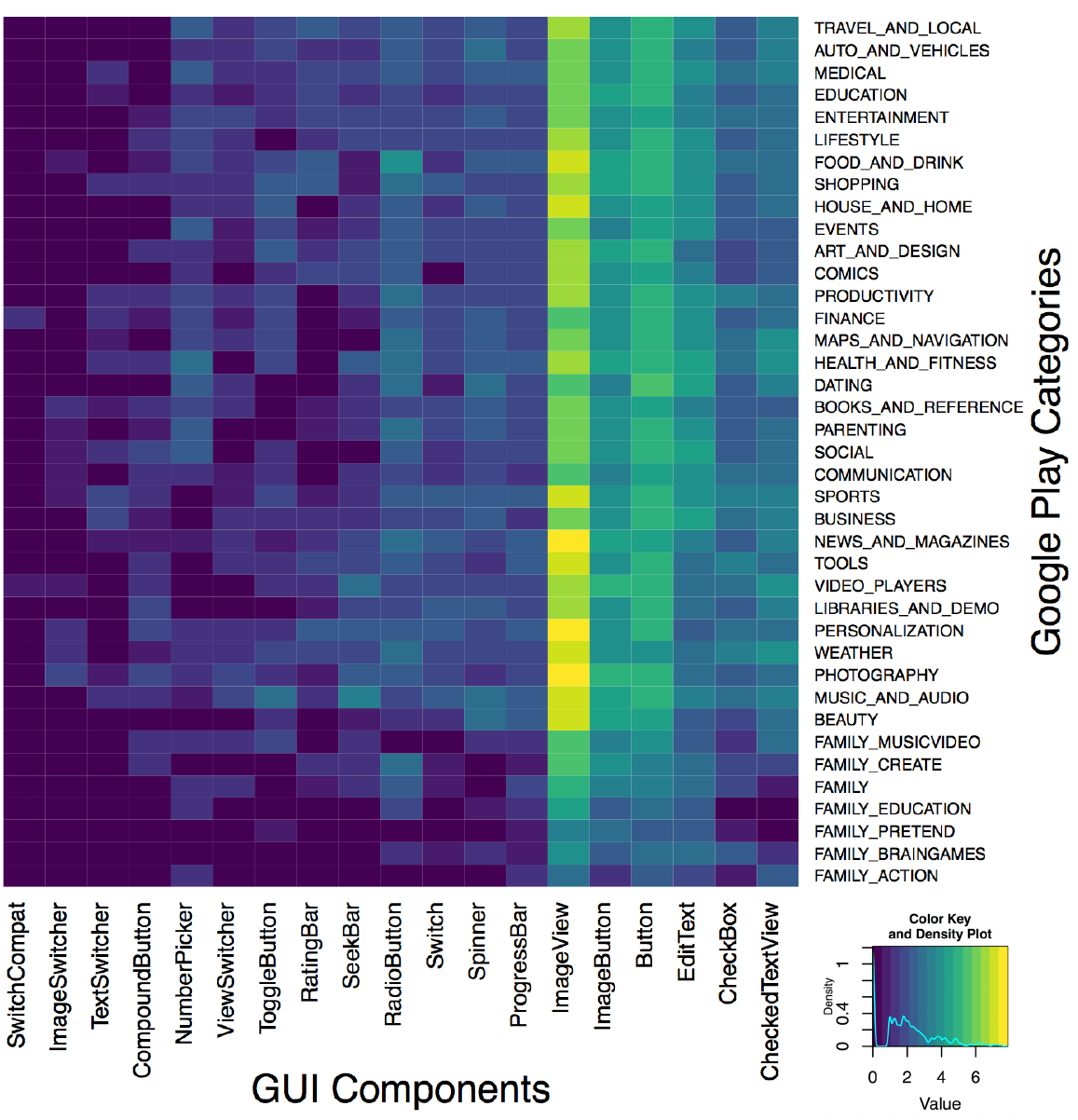}
	\caption{Heat-map of GUI Components by Category Before Filtering}
	\label{fig:components-by-category}
	\vspace{-0.3em}
\end{figure}

This process resulted in a dataset of GUI-information and screenshots for 19,786 unique app screens containing over 431,747 native Android GUI-components and containers which, to the best of the authors knowledge, \revision{is one of the largest such datasets collected to date behind the RICO dataset~\cite{Deka:UIST17}}. In Fig. \ref{fig:components-by-category} we illustrate the frequency in logarithmic-scale of the top-19 observed components by app category using a heat-map based on the frequency of components appearing from apps within a particular category (excluding \textit{TextViews} as they are, unsurprisingly, the most popular type of component observed, comprising $\approx 25\%$ of components). The distributions of components in this dataset illustrate two major points. First, while \texttt{\small ImageViews} and \texttt{\small TextViews} tend to comprise a large number of the components observed in practice, developers also heavily rely on other types of native Android components to implement key pieces of app functionality. For instance, \texttt{\small Buttons}, \texttt{\small CheckedTextViews}, and \texttt{\small RadioButtons} combined were used over 20k times across the apps in our dataset.  Second, we observed certain types of components may be more popular for different categories of apps. For instance, apps from the category of ``\texttt{\small MUSIC\_AND\_AUDIO}" tend to make much higher use of \textit{SeekBar} and \textit{ToggleButton} components to implement the expected functionalities of a media player, such as scrubbing through music and video files. These findings illustrate that for an approach to be able to effectively generate prototypes for a diverse set of mobile apps, it must be capable of correctly detecting and classifying popular types of GUI-components to support varying functionality. 

\vspace{-0.5em}
\subsubsection{Phase 2.2 - CNN Classification of GUI-Components}
\label{subsec:research-task-2.2}

Once the labeled training data set has been collected, we need to train a ML approach to extract salient features from the GUI-component images, and classify incoming images based upon these extracted features.  To accomplish this our approach leverages recent advances in CNNs. The main advantage of CNNs over other image classification approaches is that the architecture allows for automated extraction of abstract features from image data, approximation of non-linear relationships, application of the principle of data-locality, and classification in an end-to-end trainable architecture.

\vspace{-1em}
\subsubsection{ReDraw Implementation - CNN Classifier}
\label{subsec:impl-classification}

	Once the GUI-components in a target mock-up artifact have been detected using either mock-up meta-data or CV-based techniques, \ReDraw must effectively classify these components. To accomplish this \ReDraw implements a CNN capable of classifying a target image of a GUI-component into one of the 15 most-popular types of components observed in our dataset. In this subsection, we first describe the data-cleaning process used to generate the training, validation, and test datasets (examples of which are shown in Fig. \ref{fig:example-components}) before describing our CNN architecture and the training procedure we employ.

\noindent{\textbf{Data Cleaning:}} We implemented several types of preprocessing and filtering techniques to help reduce noise.  More specifically, we implemented filtering processes at three differing levels of granularity: (i) application, (ii) screen \& (iii) GUI-component level. 

While future versions of \ReDraw may support non-native apps, to provide an appropriate scope for rigorous experimentation, we have implemented \ReDraw with support for prototyping \textit{native} Android applications. Thus, once we collected the \texttt{\small xml} and screenshot files, it is important to apply filters in order to discard applications that are non-native, including games and hybrid applications. Thus, we applied the following app-level filtering methodologies:

\begin{itemize}

\item{\textbf{\textit{Hybrid Applications:}} We filtered applications that utilize Apache Cordova \cite{apache-cordova} to implement mobile apps using web-technologies such as \texttt{\small html} and \texttt{\small CSS}. To accomplish this we first decompiled the \texttt{APKs} using \texttt{\small Apktool} \cite{apktool} to get the resources used in the application. We then discarded the applications that contained a \texttt{\small www} folder with \texttt{\small html} code inside.}

\item{\textbf{\textit{Non-Standard GUI Frameworks:}} Some modern apps utilize third party graphical frameworks or libraries to create highly-customized GUIs.  While such frameworks tend to be used heavily for creating mobile games, they can also be used to create UIs for for more traditional applications.  One such popular framework is the Unity \cite{unity} game engine.  Thus, to avoid applications that utilize this engine we filtered out applications that contain the folder structure \texttt{\small com/unity3d/player} inside the code folder after decompilation with \texttt{\small Apktool}.}
	
\end{itemize}

This \revision{process resulted in the removal of 223 applications and a dataset consisting of} of 8,655 apps to which we then applied screen-level filtering.  At the Screen-level, we implemented the following pre-processing techniques:

\begin{figure}[t]
	\centering
	\vspace{-0.5em}
	\includegraphics[width=0.85\linewidth]{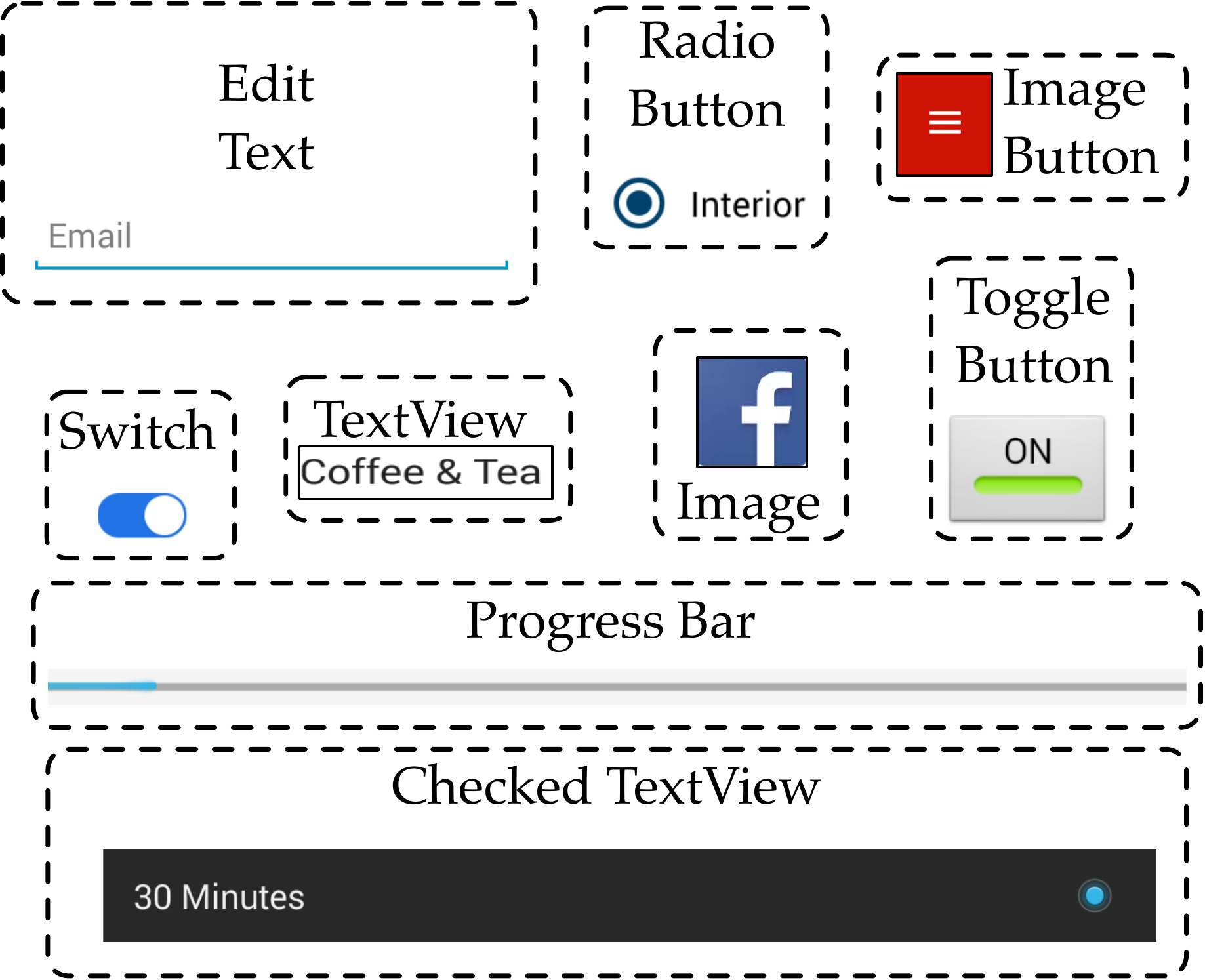}
	\caption{Example of a subset of ReDraw's training data set consisting of GUI-Component sub-images and domain (Android) specific labels. Images and corresponding Labels are grouped according to the dashed-lines.}
	\label{fig:example-components}
\end{figure}

\begin{itemize}
\item{\textbf{\textit{Filtering out Landscape screens:}} To keep the height and width of all screens consistent, we only collected data from screens that displayed in the portrait orientation.  Thus, we checked the size of the extracted screenshots and verified that the width and the height correspond to 1200x1920, the landscape oriented screen size used on our target Nexus 7 devices. However, there are some corner cases in which the images had the correct portrait size but it was on landscape. So, to overcome this we checked the extracted \texttt{\small uiautomator xml} file and validated the size of the screen to ensure a portrait orientation.}

\item{\textbf{\textit{Filtering Screens containing only Layout components:}} In Android, Layout components are used as containers that group together other types of functional components such as \textit{Buttons} and \textit{Spinners}.  However, some screens may consist only of layout components. Thus to ensure variety in our dataset, we analyzed the \texttt{\small uiautomator xml} files extracted during dynamic analysis to discard screens that are only comprised of \texttt{Layout} components such as \textit{LinearLayout, GridLayout, and FrameLayout} among others.}

\item{\textbf{\textit{Filtering WebViews:}} While many of the most popular Android apps are native, some apps may be hybrid in nature, that is utilizing web content within a native app wrapper.  Because such apps use components that cannot be extracted via \texttt{\small uiautomator} we discard them from our dataset by removing screens where a \texttt{\small WebView} occupied more than 50\% of the screen area.} 
\end{itemize}
  
\begin{figure}[t]
	\centering
	\vspace{-0.4em}
	\includegraphics[width=\linewidth]{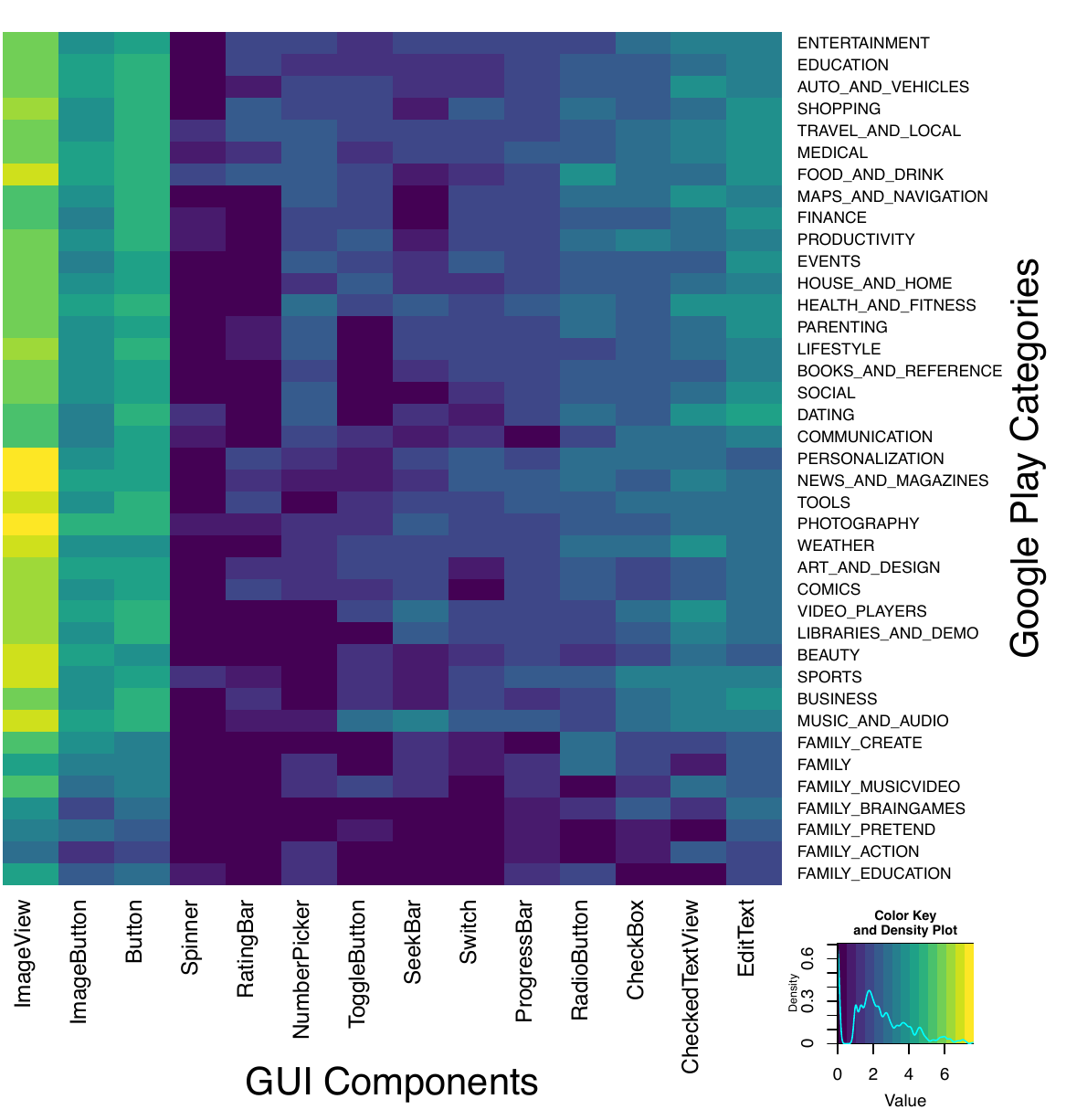}
	\caption{Heat-map of GUI Components by Category After Filtering}
	\label{fig:components-by-category-filtered}
	\vspace{-0.3em}
\end{figure}

\revision{After these filtering techniques were applied, 2,129 applications and 4,954 screens were removed, and the resulting dataset contained 14,382 unique screens with 431,747 unique components from 6,538 applications.} We used the information in the \texttt{\small uiautomator xml} files to extract the bounding boxes of \textit{leaf-level} GUI-components in the GUI-hierarchies.  We only extract leaf-level components in order to align our dataset with components detected from mock-ups. Intuitively it is unlikely that container components (\eg non-leaf nodes) would exhibit significant distinguishable features that a ML approach would be able to derive in order to perform accurate classification (hence, the use of our KNN-based approach is described in Sec. \ref{subsec:research-task-3}). Furthermore, it is unclear how such a GUI-container classification network would be used to iteratively build a GUI-structure. We performed a final filtering of the extracted leaf components: 

\begin{itemize}
\item{\textbf{\textit{Filtering Noise:}} We observed that in rare cases the bounds of components would not be valid (\eg extending beyond the borders of the screen, or represented as zero or negative areas) or components would not have a type assigned to them. Thus, we filter out these cases.}

\item{\textbf{\textit{Filtering Solid Colors:}}  We also observed that in certain circumstances, extracted components were made up of a single solid color, or in rarer cases two solid colors.  This typically occurred due to instances where the view hierarchy of a screen had loaded, but the content was still rendering on the page or being loaded over the network, when a screenshot was captured.  Thus, we discarded such cases.}

\item{\textbf{\textit{Filtering Rare GUI-Components:}} In our dataset we found that some components only appeared very few times, therefore, we filtered out any component with less than 200 instances in the initial dataset, leading to 15 GUI-component types in our dataset.}

\end{itemize}

\begin{figure}[t]
	\centering
	\includegraphics[width=0.9\linewidth]{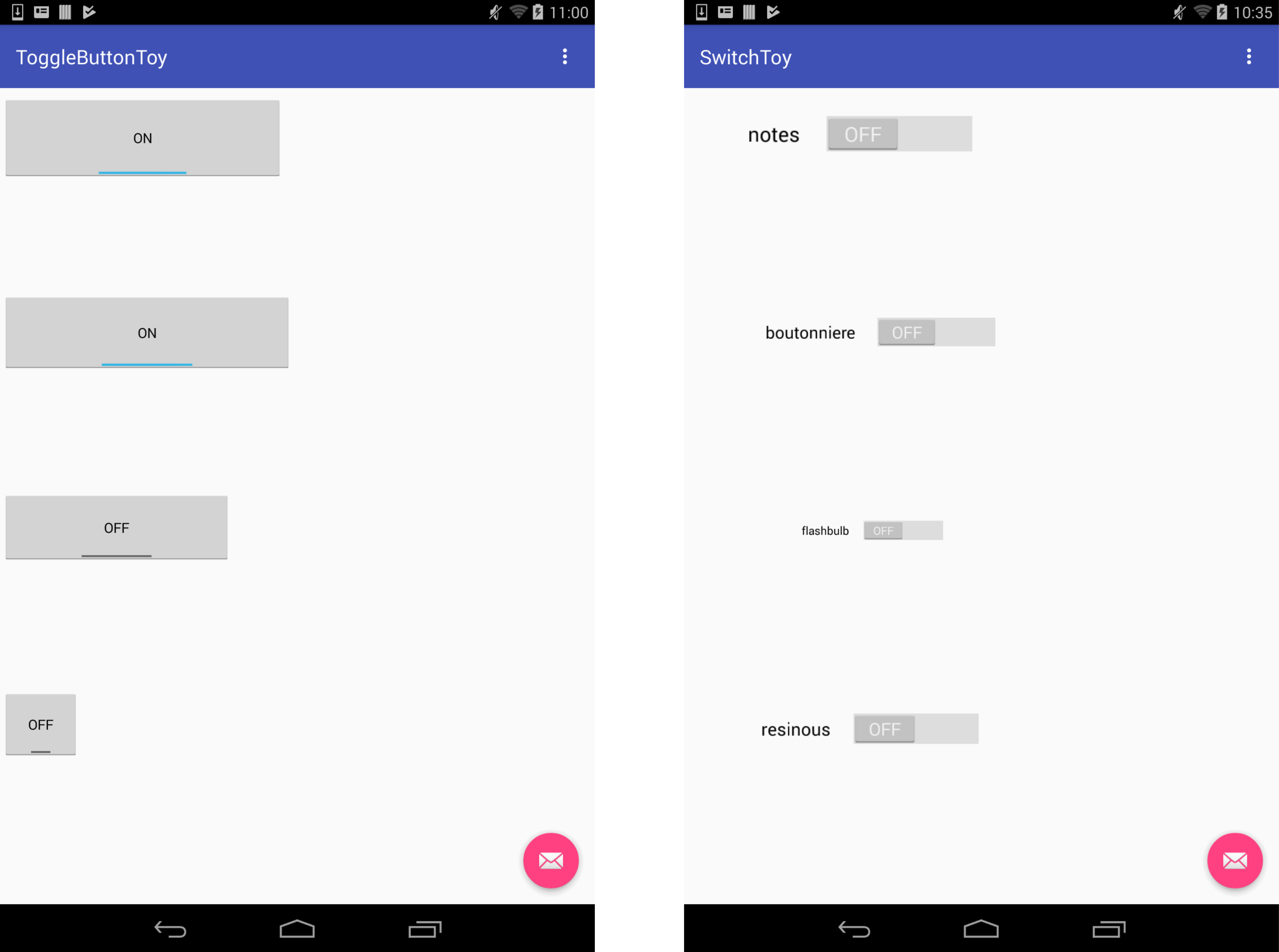}
	\caption{Screenshots of synthetically generated applications containing toggle buttons and switches}
	\vspace{-0.3em}
	\label{fig:synthetic-app-example}
\vspace{1em}
\end{figure}

\revision{The data-cleaning process described above resulted in the removal of 240,447 components resulting in 191,300 labeled images of GUI-components from 6,538 applications. We provide a heat-map illustrating the popularity of components across apps from diferent Google Play categories in Fig. \ref{fig:components-by-category-filtered}} To ensure the integrity of our dataset, we randomly sampled a statistically significant sample of 1,000 GUI-component images (corresponding to confidence interval of $\pm 3.09$ at a 95\% confidence level), and had one author manually inspect all 1,000 images and labels to ensure the dataset integrity.

\noindent{\textbf{Data Augmentation:}} Before segmenting the resulting data into training, test, and validation sets, we followed procedures from previous work \cite{Krizhevsky:NIPS12} and applied data augmentation techniques to increase the size of our dataset in order to ensure proper training support for underrepresented classes and help to combat overfitting to the training set. Like many datasets procured using ``naturally" occurring data, our dataset suffers from imbalanced classes. That is, the number of labeled images in our training set are skewed toward certain classes, resulting in certain classes that have high support, and others that have low support.  Thus, to balance our dataset, we performed two types of data augmentation: \textit{synthetic app generation} and  \textit{color perturbation}. For the sake of clarity, we will refer to data collected using our automated dynamic analysis approach as \textit{organic data} (\ie the data extracted from Google Play) and data generated via synthetic means as \textit{synthetic data} (\ie generated either via synthetic app generation or color perturbation).

	To generate synthetic data for underrepresented components, we implemented an \textit{app synthesizer} capable of generating Android apps consisting of only underrepresented components.  The app synthesizer is a Java application that is capable of automatically generating single-screen Android applications containing four instances of GUI-components (with randomized attributes) for 12 GUI-component classes in our dataset that had less than 10K observable instances.  The synthesizer places the four GUI-components of the specified type on a single app screen with randomized sizes and values (\eg numbers for a number picker, size and state for a toggle button).  Two screenshots of synthesized applications used to augment the Toggle button and Switch classes are illustrated in Fig. \ref{fig:synthetic-app-example}. We ran these apps through our \textit{Execution Engine}, collecting the \texttt{\small uiautomator xml} files and screenshots from the single generated screen for each app.  After the screenshots and uiautomator files were collected, we extracted \textit{only} the target underrepresented components from each screenshot (note that in Fig. \ref{fig:synthetic-app-example} there is a header title and button generated when creating a standard Android app), all other component types are ignored.  250 apps for each underrepresented GUI-component were synthesized, resulting in creating an extra 1K components for each class and 12K total additional GUI-components.\\

	While our application generator helps to rectify the imbalanced class support to an extent, it does not completely balance our classes and may be prone to overfitting.  Thus, to ensure proper support across all classes and to combat overfitting, we follow the guidance outlined in related work \cite{Krizhevsky:NIPS12} to perform \textit{color perturbation} on both the organic and synthetic images in our dataset.  More specifically, our color perturbation procedure extracts the RGB values for each pixel in an input image and converts the values to the HSB (Hue, Saturation, Brightness) color space.  The HSB color space represents colors as part of a cylindrical or cone model where color hues are represented by degrees.  Thus, to shift the colors of a target image, our perturbation approach randomly chooses a degree value by which each pixel in the image is shifted.  This ensures that color hues that were the same in the original image, all shift to the same new color hue in the perturbed image, preserving the visual coherency of the perturbed images.  We applied color perturbation to the training set of images until each class of GUI-component had at least 5K labeled instances, as described below.

\begin{figure}[t]
	\centering
	\includegraphics[width=0.75\linewidth]{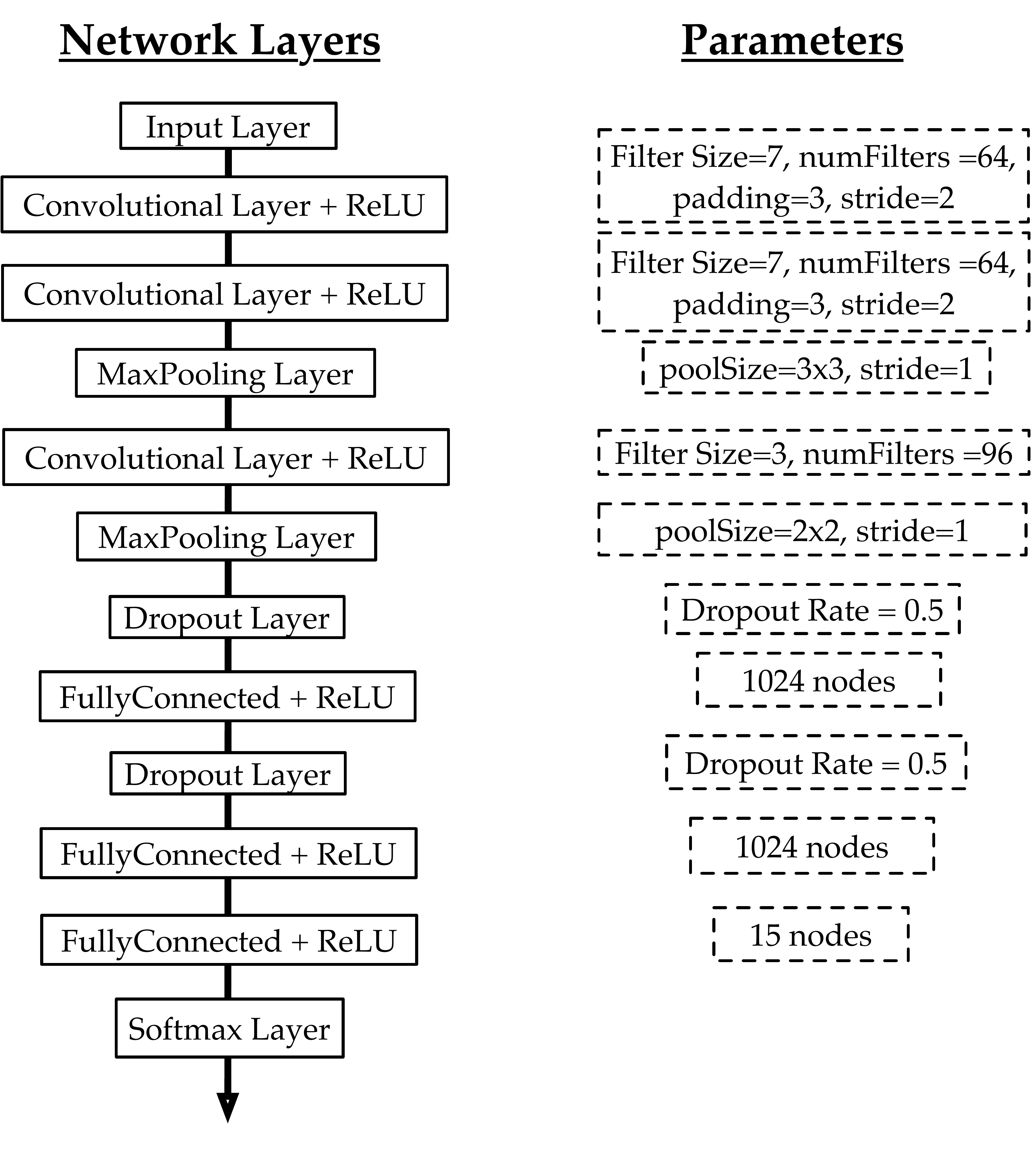}
\vspace{-2em}
	\caption{\ReDraw CNN Architecture}
	\label{fig:cnn-arch}
\end{figure}

\noindent{\textbf{Data Segmentation:}}   We created a the training, validation, and test datasets for our CNN such that the training dataset contained both \textit{organic} and \textit{synthetic} data, but the test and validation datasets contained \textbf{only} \textit{organic} data, unseen in the training phase of the CNN. To accomplish this, we randomly segmented our dataset of \textit{organic} components extracted from Google Play into \textit{training} (75\%), \textit{validation} (15\%), and \textit{test} (10\%) sets. Then for the training set, we added the synthetically generated components to the set of organic GUI-component training images, and performed color perturbation on \textbf{only} the training data (after segmentation) until each class had at least 5K training examples.  Thus, the training set contained \textit{both} organic and synthetically generated data, and the validation and test sets contained \textit{only} organic data. This segmentation methodology closely follows prior work on CNNs  \cite{Krizhevsky:NIPS12}.

\noindent{\textbf{ReDraw's CNN Architecture:}} Our CNN architecture is illustrated in Fig. \ref{fig:cnn-arch}. Our network uses an architecture similar to that of AlexNet \cite{Krizhevsky:NIPS12}, with two less convolutional layers (3 instead of 5), and is implemented in \texttt{\small MATLAB} using the Neural Network \cite{matlab-nn}, Parallel Computing \cite{matlab-pc}, and Computer Vision \cite{matlab-cv} toolkits.  While ``deeper" architectures do exist \cite{Zeiler:ECCV14,Szegedy:CVPR15,He:CVPR16} and have been shown to achieve better performance on large-scale image recognition benchmarks, this comes at the cost of dramatically longer training times and a larger set of parameters to tune. Since our goal is to classify 15 classes of the most popular Android GUI-components, we do not need the capacity of deeper networks aiming to classify thousands of image categories. We leave deeper architectures and larger numbers of image categories as future work.  Also, this allowed our CNN to converge in a matter of hours rather than weeks, and as we illustrate, still achieve high precision.  

\begin{algorithm}[t]
\caption{KNN Container Determination}
\label{alg:cont-det}
\SetAlgoLined
\footnotesize
\KwIn{InputNodes \tcp{Either leaf components or other containers}} 
\KwOut{Containers \tcp{Groupings of input components}}

\While{canGroupMoreNodes() \tcp{While groupings exist}}{
\tcp{For each screen in the mined data}
 \ForEach{$Screen\ S \in \textit{Dataset} $}
 { 
  \quad\ \ \ $TargetNodes$ = $S$.getTargetNodes()
  $\textit{score} = \frac{TargetNodes() \cap InputNodes}{TargetNodes() \cup InputNodes}$ \tcp{IOU}
  \If{$\textit{score} > \textit{curmax}$}{
  
  \textit{curmax} = \textit{score} 
  
  \textit{MatchedScreen} = $S$
  }
 }
{
 $TargetNodes$ = $MatchedScreen$.getTargetNodes()
 \textit{InputNodes}.remove($\textit{TargetNodes} \cap \textit{InputNodes})$
 \textit{Containers}.addContainers(\textit{MatchedScreen})
}
}
\end{algorithm}

To tune our CNN, we performed small scale experiments by randomly sampling 1K images from each class to build a small training/validation/test set (75\%, 15\%, 10\%) for faster training times (Note, these datasets are separate from the full set used to train/validate/test the network described earlier).  During these experiments we iteratively recorded the accuracy on our validation set, and recorded the final accuracy on the test set.  We tuned the location of layers and parameters of the network until we achieved peak test accuracy with our randomly sampled dataset.
\\

\noindent{\textbf{Training the CNN:}} To train \ReDraws network we utilized our derived training set; we trained our CNN end-to-end using back-propagation and stochastic gradient descent with momentum (SGDM), in conjunction with a technique to prevent our network from overfitting to our training data.  That is, every five epochs (\eg entire training set passing through the network once) we test the accuracy of our CNN on the validation set, saving a copy of the learned weights of the classifier at the same time.  If we observe our accuracy decrease for more than two checkpoints, we terminate the training procedure. We varied our learning rate from 0.001 to $1\times10^{-5}$ after 50 epochs, and then dropped the rate again to $1\times10^{-6}$ after 75 epochs until training terminated.  Gradually decreasing the learning rate allows for the network to ``fine-tune" the learned weights over time, leading to an increase in overall classification precision \cite{Krizhevsky:NIPS12}. Our network training time was 17 hours, 12 minutes on a machine with a single Nvidia Tesla K40 GPU.
\\

\noindent{\textbf{Using the CNN for Classification:}} Once the CNN has been trained, new, unseen images can fed into the network resulting a series of classification scores corresponding to each class.  In the case of ReDraw, the component class with the highest confidence is assigned to be the label for a given target image. We present an evaluation of the classification accuracy of \ReDraws CNN using the dataset described in this subsection later in Sec. \ref{sec:study} \& \ref{sec:results}.  

\subsection{Phase 3 - Application Assembly}
\label{subsec:research-task-3}

\begin{figure}[t]
	\centering
	\includegraphics[width=0.9\linewidth]{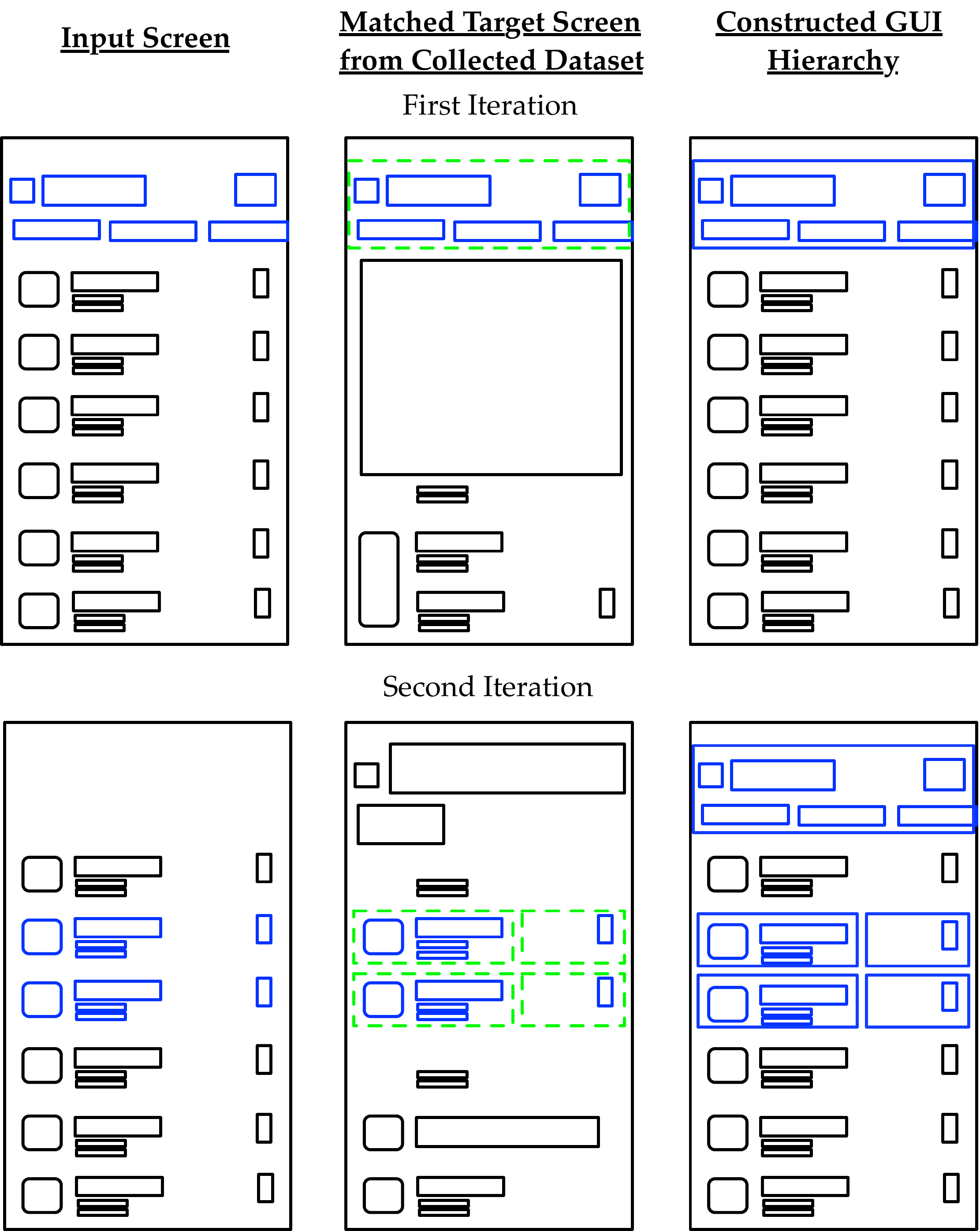}
	\caption{Illustration of  KNN Hierarchy Construction}
	\label{fig:knn-example}
\end{figure}

	The final task of the prototyping process is to assemble app GUI code, which involves three phases (Fig. \ref{fig:framework-overview}-\circled{3}): (i) building a proper hierarchy of components and containers, (ii) inferring stylistic details from a target mock-up artifact, and (iii) assembling the app.

\subsubsection{Deriving GUI-Hierarchies}
\label{subsec:knn}

	In order to infer a realistic hierarchy from the classified set of components, our approach utilizes a KNN technique (Alg. \ref{alg:cont-det}) for constructing the GUI hierarchy. This algorithm takes the set of detected and classified GUI-components represented as nodes in a single level tree ($InputNodes$) as input.  \revision{Then, for each screen in our dataset collected from automated dynamic analysis, Alg. \ref{alg:cont-det} first extracts a set of $TargetNodes$ that correspond the hierarchy level of the $InputNodes$ (Alg. \ref{alg:cont-det} -line 4), which are leaf nodes for the first pass of the algorithm. Next, the $InputNodes$ are compared to each set of extracted ($TargetNodes$) using a similarity metric based on the intersection over union (IOU) of screen area occupied by the bounding boxes of overlapping components (Alg. \ref{alg:cont-det} -line 5). A matching screen is selected by taking the screen with the highest combined IOU score between the $InputNodes$ and $TargetNodes$. Then, the parent container components from the components in the matched screen are selected as parent components to the matched $InputNodes$. The matched $InputNodes$ are then removed from the set, and the algorithm proceeds to match the remaining $InputNodes$ that were not matched during the previous iteration. This procedure is applied iteratively (including grouping containers in other containers) until a specified number of levels in the hierarchy are built or all nodes have been grouped. An illustration of this algorithm is given in Figure \ref{fig:knn-example}, where matched components are highlighted in blue and containers are represented as green boxes.

	It should be noted that all attributes of a component container are inherited during the hierarchy construction, including their type (\eg \texttt{LinearLayout}, \texttt{RelativeLayout}).  We can specify the number of component levels to ensure that hierarchies do not grow so large such that they would cause rendering delays on a device. The result of this process is a hierarchy built according to its similarity to existing GUI-hierarchies observed in data. Given different types of containers may behave differently, this technique has the advantage that, in addition to leaf level GUI-components being properly classified by the CNN, proper types of container components are built into the GUI-hierarchy via this KNN-based approach.}

\begin{figure}[t]
\begin{lstlisting}[language=Java, caption={ReDraw's Skeleton Main Activity Class\label{lst:skeleton-class}}]
public class MainActivity extends Activity {
    @Override
    protected void onCreate(Bundle savedInstanceState) {
        super.onCreate(savedInstanceState);
        setContentView(R.layout.main_activity);
    }
}
\end{lstlisting}
\end{figure}

\begin{figure}[t]
\vspace{-2em}
\begin{lstlisting}[language=Xml, caption={Snippet from \texttt{\small layout.xml} file generated by ReDraw for the Yelp Application}\label{lst:ex-code}]
<LinearLayout android:id="@+id/LinearLayout452" android:layout_height="127.80186dp" android:layout_marginStart="0.0dp" android:layout_marginTop="0.0dp" android:layout_width="400.74304dp" android:orientation="vertical" android:text="" android:textSize="8pt">
    <Button android:id="@+id/Button454" android:layout_height="58.45201dp" android:layout_marginStart="0.0dp" android:layout_marginTop="0.0dp" android:layout_width="400.74304dp" android:text="Sign up with Google" android:textSize="8pt" style="@style/Style65"/>
    <Button android:id="@+id/Button453" android:layout_height="50.526318dp" android:layout_marginStart="3.4674923dp" android:layout_marginTop="18.82353dp" android:layout_width="393.31268dp" android:text="Sign up with Facebook" android:textSize="8pt" style="@style/Style66"/>
</LinearLayout>
\end{lstlisting}
\end{figure}
\subsubsection{Inferring Styles and Assembling a Target App}
\label{subsec:inferring-styles}

To infer stylistic details from the mock-up, our approach employs the CV techniques of Color Quantization (CQ), and Color Histogram Analysis (CHA).  For GUI-components whose type does not suggest that they are displaying an image, our approach quantizes the color values of each pixel and constructs a color histogram. The most popular color values can then be used to inform style attributes of components when code is generated. For example, for a component displaying text, the most prevalent color can be used as a background and the second most prevalent color can be used for the font.

\begin{figure}[t]
\vspace{-1em}
\hspace*{1cm}
\begin{lstlisting}[language=Xml, caption={Snippet from \texttt{\small style.xml} file generated by ReDraw for the Yelp Application}\label{lst:ex-style}]
<style name="Style63" parent="AppTheme">
   <item name="android:textColor">#FEFEFF</item>
</style>
<style name="Style64" parent="AppTheme">
   <item name="android:textColor">#FEFEFF</item>
</style>
<style name="Style65" parent="AppTheme">
   <item name="android:background">#DD4B39</item>
   <item name="android:textColor">#FEFEFF</item>
</style>
\end{lstlisting}
\end{figure}

\subsubsection{ReDraw Implementation - App Assembly}
\label{subsec:impl-assembly}

ReDraw assembles Android applications, using the KNN approach for GUI-hierarchy construction (see Sec. \ref{subsec:research-task-3}.1) and CV-based detection of color styles.  The input to Alg. \ref{alg:cont-det} is the set of classified ``leaf-node" components from the CNN, and the output is a GUI-hierarchy. To provide sufficient data for the KNN-algorithm, a corpus including all of the info from the "cleaned" screens of the GUI-hierarchies mined from our large scale dynamic analysis process is constructed. This corpus forms the dataset \textit{TargetNodes} to which the \textit{InputNode} components are matched against during hierarchy construction.  The GUI-hierarchy generated by the KNN for the target "leaf-node" components is then used to infer stylistic details from the original mock-up artifact.  More specifically, for each component and container, we perform CQ and CHA to extract the dominant colors for each component. For components which have a text element, we apply optical character recognition (OCR) using the open source Tesseract \cite{tesseract-ocr} library on the original screenshot to obtain the strings.

\revision{Currently, our approach is able to infer three major types of stylistic detail from target components:}

\begin{itemize}
\revision{
	\item{\textit{\textbf{Background Color:}} To infer the background color of components and containers, ReDraw simply utilizes the dominant color in the CHA for a specific component as the background color.}
	\item{\textit{\textbf{Font Color:}} To infer the font color for components, ReDraw uses the dominant color in the CHA as the background text and the second most dominant color as the font color.}
	\item{\textit{\textbf{Font Size:}} ReDraw is able to infer the font size of textual components by using the pixel based height of the bounding boxes of text-related components.}
}
\end{itemize}

\revision{These techniques are used for both variants of the \ReDraw approach (\eg mock-up based and CV based). There is ample opportunity for future work to improve upon the inference of stylistic  details, particularly from mock-up artifacts. More specifically, future work could expand this process to further adapt the style of ``standard'' components to match stylistic details observed in a mock-up artifact. Depending upon the export format for a mock-up, ReDraw could also potentially infer additional styles such as the font utilized or properties of component shapes (\eg button bevels). While \ReDraws current capabilities for inferring stylistic details are limited to the above three categories, in Section \ref{sec:results} we illustrate that these are sufficient to enable \ReDraw to generate highly visually similar applications in comparison to target images.}

	\ReDraw encodes the information regarding the GUI-hierarchy, stylistic details, and strings detected using OCR into an intermediate representation (IR) before translating it into code.  This IR follows the format of \texttt{\small uiautomator} \texttt{\small xml} files that describes dynamic information from an Android screen.  Thus, after \ReDraw encodes the GUI information into the \texttt{\small uiautomator}-based IR, it then generates the necessary resource \texttt{\small xml} files (\eg files in the \texttt{\small res} folder of an Android app project directory) by parsing the \texttt{\small uiautomator}-based IR \texttt{\small xml} file.   This process generates the following two types of resource code for the generated app: (i) the \texttt{\small layout.xml} code describing the general GUI structure complete with strings detected via OCR; and (ii) a \texttt{\small style.xml} file that stipulates the color and style information for each component gleaned via the CV techniques, and ReDraw generates the xml source files following the best practices stipulated in the Android developer guidelines \cite{android-ui-development}, such as utilizing relative positioning, and proper padding and margins. In addition to these resource \texttt{\small xml} files \ReDraw also generates a skeleton Java class encompassing the \texttt{\small MainActivity} which renders the GUI stipulated in the resource \texttt{\small xml} files, as well as other various files required to build and package the code into an \texttt{\small apk}. The Skeleton \texttt{\small MainActivity} Java class is shown in Listing \ref{lst:skeleton-class} and snippets from generated \texttt{\small layout.xml} \& \texttt{\small style.xml} files for a screen from the Yelp application are shown in Listings \ref{lst:ex-code} \& \ref{lst:ex-style}.  The \texttt{\small layout.xml} snippet of code generated by ReDraw illustrates the use of margins and relative \texttt{\small dp} values to stipulate the spatial properties of GUI-containers and GUI-components and references the \texttt{\small style.xml} file to stipulate color information. Listing \ref{lst:ex-style} illustrates the corresponding styles and colors referenced by the \texttt{\small layout.xml} file.


\vspace{-1em}
\section{Empirical Study Design}
\label{sec:study}

The \textit{goal} of our empirical study is to evaluate \ReDraw in terms of (i) the accuracy of the CNN GUI-component classifier, (ii) the similarity of the generated GUI-hierarchies to real hierarchies constructed by developers, (iii) the visual similarity of generated apps compared to mock-ups, and (iv) ReDraw's suitability in an industrial context. The \textit{context} of this study consists of (i) a set of 191,300 labeled images of Android GUI-components extracted from 14,382 unique app screens mined from 6,538 \texttt{APKs} from the Google Play store (see Sec. \ref{subsubsec:impl-app-mining} for details) to assess the accuracy of the CNN-classifier, (ii) 83 additional screens (not included in the dataset to train and test the CNN-classifier) extracted from 32 of the highest rated apps on Google Play (top-3 in each category), (iii) nine reverse engineered Sketch mockups from eight randomly selected highly rated Google Play Apps to serve as mock-up artifacts, and (iv) two additional approaches for prototyping Android applications \Remaui \cite{Nguyen:ASE15} and pix2code \cite{Beltramelli:arXiv17}. The \textit{quality focus} of this study is the effectiveness of \ReDraw to generate prototype apps that are both visually similar to target mock-up artifacts, with GUI-hierarchies similar to those created by developers. To aid in achieving the goals of our study we formulated the following RQs:

\begin{itemize}
	\item{ \textit{\textbf{RQ$_1$}: How accurate is the CNN-based image classifier for classifying Android GUI-components?}}
	\item{ \textit{\textbf{RQ$_2$}: How similar are GUI-hierarchies constructed using \ReDraws KNN algorithm compared to real GUI-hierarchies?}}
	\item{ \textit{\textbf{RQ$_3$}: Are the prototype applications that \ReDraw generates visually similar to mock-up artifacts?}}
	\item{ \textit{\textbf{RQ$_4$}: Would actual mobile developers and designers consider using \ReDraw as part of their workflow?}}
\end{itemize} 

\revision{It should be noted that in answering RQ$_2$-RQ$_4$ we use two types of mock-up artifacts (existing application screenshots, and reverse engineered Sketch mock-ups) as a proxy  for real GUI-design mock-ups, and these artifacts are  not a perfect approximation. More specifically, screenshots represent a finalized GUI-design, whereas real GUI design mockups may not be complete and might include ambiguities or design parameters that are able to be properly implemented in code (\ie unavailable fonts or impractical spatial layouts). Thus, we do not claim to measure \ReDraws performance on incomplete or ``in-progress'' design mock-ups. However, it was not possible to obtain actual GUI design mock-ups for our study, and our target screenshots and reverse engineered mock-ups stem from widely used applications. We discuss this point further in Sec. \ref{sec:limitations-threats}}.

\vspace{-1em}
\subsection{RQ$_1$: Effectiveness of the CNN}
\label{subsec:rq1}

\begin{table}[]
\centering
\caption{Labeled GUI-Component Image Datasets}
\vspace{-0.5em}
\label{tab:datasets}
\scriptsize
\begin{tabular}{|l|l|l|l|l|l|}
\hline
\textbf{GUI-C Type} & \textbf{Total \# (C)} & \textbf{Tr (O)} & \textbf{Tr (O+S)} & \textbf{Valid} & \textbf{Test} \\ \hline
TextView                    & 99,200                                & 74,087                           & 74,087                                       & 15,236                   & 9,877              \\ \hline
ImageView                   & 53,324                                & 39,983                           & 39,983                                       & 7,996                    & 5,345              \\ \hline
Button                      & 16,007                                & 12,007                           & 12,007                                       & 2,400                    & 1,600              \\ \hline
ImageButton                 & 8,693                                 & 6,521                            & 6,521                                        & 1,306                    & 866               \\ \hline
EditText                    & 5,643                                 & 4,230                            & 5,000                                        & 846                     & 567               \\ \hline
CheckedTextView             & 3,424                                 & 2,582                            & 5,000                                        & 505                     & 337               \\ \hline
CheckBox                    & 1,650                                 & 1,238                            & 5,000                                        & 247                     & 165               \\ \hline
RadioButton                 & 1,293                                 & 970                             & 5,000                                        & 194                     & 129               \\ \hline
ProgressBar                 & 406                                  & 307                             & 5,000                                        & 60                      & 39                \\ \hline
SeekBar                     & 405                                  & 304                             & 5,000                                        & 61                      & 40                \\ \hline
NumberPicker                & 378                                  & 283                             & 5,000                                        & 57                      & 38                \\ \hline
Switch                      & 373                                  & 280                             & 5,000                                        & 56                      & 37                \\ \hline
ToggleButton                & 265                                  & 199                             & 5,000                                        & 40                      & 26                \\ \hline
RatingBar                   & 219                                  & 164                             & 5,000                                        & 33                      & 22                \\ \hline
Spinner                     & 20                                   & 15                              & 5,000                                        & 3                       & 2                 \\ \hline
\textbf{Total}              & 191,300                               & 143,170                          & 187,598                                      & 29,040                   & 19,090             \\ \hline
\end{tabular}
\vspace{0.3em}
\caption*{\footnotesize \textbf{ Abbreviations for column headings:} ``Total\#(C)"=Total \# of GUI-components in each class after cleaning; ``Valid"= Validation; ``Tr(O)"= Training Data (Organic Components Only); ``Tr(O+S)"= Training Data (Organic + Synthetic Components).}
\vspace{-1.5em}
\end{table}
{\normalsize
	To answer \textbf{RQ$_1$}, as outlined in Sec. \ref{subsec:impl-classification} we applied a large scale automated dynamic analysis technique and various data cleaning procedures which resulted in a total of 6,538 apps, 14,382 unique screens, and 191,300 labeled images of GUI-components.  To normalize support across classes and prepare training, validation and test sets in order measure the effectiveness of our CNN we applied data augmentation, and segmentation techniques also described in detail in Sec. \ref{subsec:impl-classification}.  The datasets utilized are illustrated, broken down by class, in Table \ref{tab:datasets}.  We trained the CNN on the training set of data, avoiding overfitting using a validation set as described in Sec. \ref{subsec:impl-classification}. To reiterate, all of the images in the test and validation sets were extracted from real applications and were separate (\eg unseen) from the training set. To evaluate the effectiveness of our approach we measure the average top-1 classification precision across all classes on the Test set of data:}

\begin{small}
\vspace{-0.5em}
$$ P = \frac{TP}{TP + FP}$$
\end{small}
\noindent 
where $TP$ corresponds to true positives, or instances where the top class predicted by the network is correct, and $FP$ corresponds to false positives, or instances where the top classification prediction of the network is not correct.  To illustrate the classification capabilities of our CNN, we present a confusion matrix with precision across classes in Sec. \ref{sec:results}.  The confusion matrix illustrates correct true positives across the highlighted diagonal, and false positives in the other cells. To help justify the need and applicability of a CNN-based approach, we measure the classification performance of our CNN against a baseline technique, as recent work has suggested that deep learning techniques applied to SE tasks should be compared to simpler, less computationally expensive alternatives \cite{Fu:FSE17}. To this end, we implemented a baseline Support Vector Machine (SVM) for classification based image classification approach \cite{Csurka:04} that utilizes a "Bag of Visual Words" (BOVW). At a high level, this approach extracts image features using the Speeded-Up Robust Feature (SURF) detection algorithm \cite{Bay:CVIU08}, then uses K-means clustering to cluster similar features together, and utilizes an SVM trained on resulting feature clusters. We utilized the same training/validation/test set of data used to the train the CNN and followed the methodology in \cite{Csurka:04} to vary the number of K-means clusters from $k=1,000$ to $k=5,000$ in steps of $50$, finding that $k=4,250$ achieved the best performance in terms of classification precision for our dataset. We also report the confusion matrix of precision values for the BOVW technique.

\subsection{RQ$_2$: GUI Hierarchy Construction}
\label{subsec:rq2}

	In order to answer \textbf{RQ$_2$} we aim to measure the similarity of the GUI-hierarchies in apps generated by \ReDraw compared to a ground truth set of hierarchies and a set of hierarchies generated by two baseline mobile app prototyping approaches, \Remaui and pix2code. To carry out this portion of the study, we selected 32 apps from our cleaned dataset of \texttt{\small Apks} by randomly selecting one of the top-10 apps from each category (grouping all ``Family" categories together). We then manually extracted 2-3 screenshots and \texttt{uiautomator xml} files per app, which were not included in the original dataset used to train, validate or test the CNN. After discarding screens according to our filtering techniques, this resulted in a set of 83 screens.  Each of these screens was used as input to \ReDraw, \Remaui, and pix2code from which a prototype application was generated.  Ideally, a comparison would compare the GUI-related source code of applications (\eg \texttt{\small xml} files located in the \texttt{\small res} folder of Android project) generated using various automated techniques however, the source code of many of the subject Google Play applications is not available.  Therefore, to compare GUI-hierarchies, we compare the runtime GUI-hierarchies extracted dynamically from the generated prototype apps for each approach using \texttt{\small uiautomator}, to the set of ``ground truth" \texttt{\small uiautomator xml} files extracted from the original applications.  The \texttt{\small uiautomator} representation of the GUI is a reflection of the automatically generated GUI-related source code for each studied prototyping approach displayed at runtime on the device screen. This allows us to make an accurate comparison of the hierarchal representation of GUI-components and GUI-containers for each approach.

	To provide a performance comparison to \ReDraw, we selected the two most closely related approaches in related research literature, \Remaui \cite{Nguyen:ASE15} and \pixcode \cite{Beltramelli:arXiv17} , to provide a comparative baseline. To provide a comparison against \pixcode, we utilized the code provided by the authors of the paper on GitHub \cite{pix2code-github} and the provided training dataset of synthesized applications.  We were not able to train the \pixcode approach on our mined dataset of Android application screenshots for two reasons: (i) \pixcode uses a proprietary domain specific language (DSL) that training examples must be translated to and the authors do not provide transformation code or specifications for the DSL, (ii) the \pixcode approach requires the GUI-related source code of the applications for training, which would have needed to be reverse engineered from the Android apps in our dataset from Google Play. To provide a comparison against REMAUI \cite{Nguyen:ASE15}, we re-implemented the approach based on the details provided in the paper, as the tool was not available as of the time of writing this paper\footnote{\Remaui is partially available as a web-service \cite{remaui-web}, but it did not work reliably and we could not generate apps using this interface.}.

	As stated in Sec. \ref{subsec:research-task-1} \ReDraw enables two different methodologies for for \textit{detecting} GUI-components from a mock-up artifact: (i) CV-based techniques and (ii) parsing information directly from mock-up artifacts.  We consider both of these variants in our evaluation which we will refer to as \ReDraw-CV (for the CV-based approach) and \ReDraw-Mockup (for the approach that parses mock-up metadata).  Our set of 83 screens extracted from Google Play does not contain traditional mock-up artifacts that would arise as part of the app design and development process (\eg Photoshop or Sketch files) and reverse engineering these artifacts is an extremely time-consuming task (see Sec. \ref{subsec:rq4}).  Thus, because manually reverse-engineering mock-ups from 83 screens is not practical, \ReDraw-Mockup was modified to parse \textit{only} the bounding-box information of leaf node GUI-components from \texttt{uiautomator} files as a substitute for mock-up metadata.

	We compared the runtime hierarchies of all generated apps to the original, ground truth runtime hierarchies (extracted from the original \texttt{\small uiautomator xml} files)  by deconstructing the trees using pre-order and using the Wagner-Fischer \cite{wagner-fischer} implementation of Levenshtein edit distance for calculating similarity between the hierarchical (\ie tree) representations of the runtime GUIs.  The hierarchies were deconstructed such that the \textit{type} and \textit{nested order} of components are included in the hierarchy deconstruction.  We implemented the pre-order traversal in this way to avoid small deviations in other attributes included in the \texttt{\small uiautomator} information, such as pixel values, given that the main \textit{goal} of this evaluation is to measure hierarchical similarities.

	In our measurement of edit distance, we consider three different types of traditional edit operations: insertion, deletion, and substitution. In order to more completely measure the similarity of the prototype app hierarchies to the ground truth hierarchies, we introduced a weighting schema representing a ``penalty" for each type of edit operation, wherein the default case each operation carries an identical weight of $1/3$.  We vary the weights of each edit and calculate a distribution of edit distances which are dependent on the fraction of the total penalty that a given operation (\ie insertion, deletion, or substitution) occupies, and carry out these calculations varying each operation separately. The operations that are not under examination split the difference of the remaining weight of the total penalty equally. For example, when insertions are given a penalty of 0.5, the penalties for deletion and substitution are set to 0.25 each. This helps to better visualize the minimum edit distance required to transform a \ReDraw, pix2code, or \Remaui generated hierarchy to the original hierarchy and also helps to to better describe the nature of the inaccuracies of the hierarchies generated by each method.

\subsection{RQ$_3$: Visual Similarity}
\label{subsec:rq3}

One of \ReDraws goals  is to generate apps that are visually similar to target mock-ups. Thus to answer RQ$_3$, we compared the visual similarity of apps generated by \ReDraw, pix2code, and \Remaui, using the same set of 83 apps from \textbf{RQ$_2$}.  The subjects of comparison for this section of the study were screenshots collected from the prototype applications generated by \ReDraw-CV, \ReDraw-Mockup, pix2code, and \Remaui. Following the experimental settings used to validate \Remaui \cite{Nguyen:ASE15}, we used the open source PhotoHawk \cite{Photohawk} library to measure the \textit{mean squared error} (MSE) and \textit{mean average error} (MAE) of screenshots from the generated prototype apps from each approach compared to the original app screenshots. To examine whether the MAE and MSE varied to a statistically significant degree between approaches, we compare the MAE \& MSE distributions for each possible pair of approaches using a two-tailed Mann-Whitney test \cite{Conover:1998} ($p$-value).  Results are declared as statistically significant at a $0.05$ significance level. We also estimate the magnitude of the observed differences using the Cliff's Delta ($d$), which allows for a nonparametric effect size measure for ordinal data \cite{Grissom:2005}.

\vspace{-1em}
\subsection{RQ$_4$: Industrial Applicability}
\label{subsec:rq4}

\begin{table}[]
\centering
\footnotesize
\caption{Semi-Structured Interview Questions for Developers \& Designers}
\label{tab:interview-qs}
\begin{tabular}{|p{0.05\linewidth}|p{0.80\linewidth}|}
\hline
\textbf{Q\#} & \multicolumn{1}{c|}{\textbf{Question Text}}                                                                                                                                                                                                                                                                   \\ \hline
\textbf{Q1}  & Given the scenario where you are creating a new user interface, would you consider adopting ReDraw in your design or development workflow? Please elaborate. \\ \hline
\textbf{Q2}  & What do you think of the visual similarity of the ReDraw applications compared to the original applications? Please elaborate.                                                                                                                                                                       \\ \hline
\textbf{Q3}  & Do you think that the GUI-hierarchies (\eg groupings of components) generated by ReDraw are effective? Please elaborate.                                                                                                                                                                            \\ \hline
\textbf{Q4}  & What improvements to ReDraw would further aid the mobile application prototyping process at your company? Please elaborate.                                                                                                                                                                          \\ \hline
\end{tabular}
\end{table}

Ultimately, the goal of \ReDraw is integration into real application development workflows, thus as part of our evaluation, we aim to investigate \ReDraw's applicability in such contexts. To investigate \textbf{RQ$_4$} we conducted semi-structured interviews with a front-end Android developer at Google, an Android UI designer from Huawei, and a mobile researcher from Facebook.  For each of these three participants, we randomly selected nine screens from the set of apps used in \textbf{RQ$_2$-RQ$_3$} and manually reversed engineered  Sketch mock-ups of these apps. We verified the visual fidelity of these mock-ups using the GVT tool \cite{Moran:ICSE18}, which has been used in prior work to detect presentation failures, ensuring that there were no reported design violations reported in the reverse-engineered mockups.  This process of reverse-engineering the mock-ups was extremely time-consuming to reach acceptable levels, with well over ten hours invested into each of the nine mock-ups.  We then used \ReDraw to generate apps using both CV-based detection and utilizing data from the mock-ups. Before the interviews, we sent participants a package containing the ReDraw generated apps, complete with screenshots and source code, and the original app screenshots and Sketch mock-ups.  We then asked a series of questions (delineated in Table \ref{tab:interview-qs}) related to (i) the potential applicability of the tool in their design/development workflows, (ii) aspects of the tool they appreciated, and (iii) areas for improvement.  Our investigation into this research question is meant to provide insight into the applicability of \ReDraw to fit into real design development workflows, however, we leave full-scale user studies and trials as future work with industrial collaborators. This study is not meant to be comparative, but rather to help gauge \ReDraws industrial applicability.


\vspace{-0.3cm}
\section{Experimental Results}
\label{sec:results} 

\begin{table*}[tb]
	\footnotesize
	\centering
	\caption{Confusion Matrix for \ReDraw}
	\vspace{-0.5em}
	\label{tab:confusion_matrix_redraw}
	\begin{tabular}{lr|r|r|r|r|r|r|r|r|r|r|r|r|r|r|r} \hline
		&Total&TV&IV&Bt&S&ET&IBt&CTV&PB&RB&TB&CB&Sp&SB&NP&RBt\\
		\hline
		TV&9877&\cellcolor{blue!20}94\%&3\%&2\%&0\%&0\%&0\%&0\%&0\%&0\%&0\%&0\%&0\%&0\%&0\%&0\%\\
		IV&5345&5\%&\cellcolor{blue!20}93\%&1\%&0\%&0\%&1\%&0\%&0\%&0\%&0\%&0\%&0\%&0\%&0\%&0\%\\
		Bt&1600&11\%&6\%&\cellcolor{blue!20}81\%&0\%&1\%&1\%&0\%&0\%&0\%&0\%&0\%&0\%&0\%&0\%&0\%\\
		S&37&5\%&3\%&0\%&\cellcolor{blue!20}87\%&0\%&0\%&5\%&0\%&0\%&0\%&0\%&0\%&0\%&0\%&0\%\\
		ET&567&14\%&3\%&2\%&0\%&\cellcolor{blue!20}81\%&0\%&0\%&0\%&0\%&0\%&0\%&0\%&0\%&0\%&0\%\\
		IBt&866&4\%&23\%&1\%&0\%&0\%&\cellcolor{blue!20}72\%&0\%&0\%&0\%&0\%&0\%&0\%&0\%&0\%&0\%\\
		CTV&337&7\%&0\%&0\%&0\%&0\%&0\%&\cellcolor{blue!20}93\%&0\%&0\%&0\%&0\%&0\%&0\%&0\%&0\%\\
		PB&41&15\%&29\%&0\%&0\%&0\%&0\%&0\%&\cellcolor{blue!20}56\%&0\%&0\%&0\%&0\%&0\%&0\%&0\%\\
		RB&22&0\%&0\%&0\%&0\%&0\%&0\%&0\%&0\%&\cellcolor{blue!20}100\%&0\%&0\%&0\%&0\%&0\%&0\%\\
		TBt&26&19\%&22\%&7\%&0\%&0\%&0\%&0\%&0\%&0\%&\cellcolor{blue!20}52\%&0\%&0\%&0\%&0\%&0\%\\
		CB&165&12\%&7\%&0\%&0\%&1\%&0\%&0\%&0\%&0\%&0\%&\cellcolor{blue!20}81\%&0\%&0\%&0\%&0\%\\
		Sp&2&0\%&0\%&0\%&0\%&0\%&0\%&0\%&0\%&0\%&0\%&0\%&\cellcolor{blue!20}100\%&0\%&0\%&0\%\\
		SB&39&10\%&13\%&0\%&0\%&0\%&0\%&0\%&0\%&0\%&0\%&0\%&0\%&\cellcolor{blue!20}78\%&0\%&0\%\\
		NP&40&0\%&5\%&0\%&0\%&0\%&0\%&0\%&0\%&0\%&0\%&0\%&0\%&0\%&\cellcolor{blue!20}95\%&0\%\\
		RBt&129&4\%&3\%&2\%&0\%&0\%&0\%&1\%&0\%&0\%&0\%&1\%&0\%&0\%&0\%&\cellcolor{blue!20}89\%\\
		
		\hline
	\end{tabular}
\end{table*}

\begin{table*}[tb]
	\footnotesize
	\centering
	\vspace{-0.5em}
	\caption{Confusion Matrix for BOVW Baseline}
	\vspace{-0.5em}
	\label{tab:confusion_matrix_baseline}
	\begin{tabular}{lr|r|r|r|r|r|r|r|r|r|r|r|r|r|r|r} \hline
		 &Total&TV&IV&Bt&S&ET&IBt&CTV&PB&RB&TB&CB&Sp&SB&NP&RBt\\
		\hline
		TV&9877&\cellcolor{blue!20}59\%&4\%&9\%&1\%&6\%&2\%&8\%&6\%&0\%&1\%&2\%&0\%&1\%&0\%&2\%\\
		IV&5345&4\%&\cellcolor{blue!20}51\%&4\%&1\%&2\%&11\%&2\%&18\%&1\%&1\%&3\%&0\%&2\%&0\%&2\%\\
		Bt&1600&6\%&6\%&\cellcolor{blue!20}59\%&1\%&5\%&4\%&7\%&4\%&0\%&1\%&1\%&0\%&0\%&3\%&1\%\\
		S&37&5\%&0\%&3\%&\cellcolor{blue!20}65\%&0\%&0\%&5\%&22\%&0\%&0\%&0\%&0\%&0\%&0\%&0\%\\
		ET&567&6\%&2\%&4\%&1\%&\cellcolor{blue!20}62\%&1\%&4\%&15\%&0\%&0\%&1\%&0\%&0\%&4\%&1\%\\
		IBt&866&2\%&16\%&3\%&0\%&2\%&\cellcolor{blue!20}61\%&1\%&9\%&1\%&1\%&2\%&0\%&2\%&0\%&3\%\\
		CTV&337&3\%&1\%&7\%&1\%&3\%&0\%&\cellcolor{blue!20}81\%&1\%&0\%&0\%&2\%&0\%&0\%&0\%&2\%\\
		PB&41&0\%&24\%&2\%&0\%&2\%&5\%&2\%&\cellcolor{blue!20}54\%&0\%&0\%&2\%&2\%&2\%&0\%&2\%\\
		RB&22&0\%&5\%&0\%&0\%&0\%&0\%&0\%&27\%&\cellcolor{blue!20}68\%&0\%&0\%&0\%&0\%&0\%&0\%\\
		TBt&26&7\%&7\%&19\%&0\%&0\%&0\%&11\%&15\%&0\%&\cellcolor{blue!20}33\%&0\%&0\%&0\%&0\%&7\%\\
		CB&165&4\%&2\%&3\%&1\%&2\%&1\%&2\%&12\%&1\%&0\%&\cellcolor{blue!20}72\%&0\%&0\%&0\%&1\%\\
		Sp&2&0\%&0\%&0\%&0\%&0\%&0\%&0\%&0\%&0\%&0\%&0\%&\cellcolor{blue!20}100\%&0\%&0\%&0\%\\
		SB&39&0\%&5\%&0\%&0\%&0\%&0\%&0\%&18\%&3\%&0\%&5\%&0\%&\cellcolor{blue!20}68\%&0\%&3\%\\
		NP&40&3\%&0\%&5\%&0\%&3\%&0\%&5\%&0\%&0\%&0\%&0\%&0\%&0\%&\cellcolor{blue!20}84\%&0\%\\
		RBt&129&6\%&3\%&5\%&1\%&3\%&0\%&6\%&18\%&0\%&1\%&1\%&0\%&1\%&0\%&\cellcolor{blue!20}55\%\\
		
		\hline
	\end{tabular}
\vspace{0.5em}
\caption*{Abbreviations for column headings representing GUI-component types: TextView (TV), ImageView (IV), Button (Bt), Switch (S), EditText (ET), ImageButton (IBt), CheckedTextView (CTV), ProgressBar (PB), RadioButton (RB), ToggleButton (TBt), CheckBox (CB), Spinner (Sp), SeekBar (SB), NumberPicker (NP), RadioButton (RBt)}
\vspace{-3em}
\end{table*}

\subsection{RQ$_1$ Results: Effectiveness of the CNN}
\label{subsec:results-rq1}

	The confusion matrices illustrating the classification precision across the 15 Android component classes for both the CNN-classifier and the Baseline BOVW approach are shown in Tables \ref{tab:confusion_matrix_redraw} \& \ref{tab:confusion_matrix_baseline} respectively.  The first column of the matrices illustrate the number of components in the test set, and the numbers in the matrix correspond to the percentage of each class on the y-axis, that were classified as components on the x-axis.  Thus, the diagonal of the matrices (highlighted in blue) corresponds to correct classifications. The overall top-1 precision for the CNN (based on raw numbers of components classified) is $91.1\%$, whereas for the BOVW approach the overall top-1 precision is $64.7\%$. Hence, it is clear that the CNN-based classifier that \ReDraw employs outperforms the baseline, illustrating the advantage of the CNN architecture compared to a heuristic-based feature extraction approach. In fact, \ReDraws CNN outperforms the baseline in classification precision \textit{across all classes}. 

	It should be noted that \ReDraws classification precision does suffer for certain classes, namely \texttt{\small ProgressBars} and \texttt{\small ToggleButtons}.  We found that the classification accuracy of these component types was hindered due to multiple existing styles of the components.  For instance, the \texttt{\small ProgressBar} had two primary styles, traditional progress bars, which are short in the y-direction and long in the x-direction, and square progress bars that rendered a progress wheel.  With two very distinct shapes, it was difficult for our CNN to distinguish between the drastically different images and learn a coherent set of features to differentiate the two.  While the CNN may occasionally misclassify components, the confusion matrix illustrates that these misclassifications are typically skewed toward \textit{similar} classes.  For example, \texttt{\small ImageButtons} are primarily misclassified as \texttt{\small ImageViews}, and \texttt{\small EditTexts} are misclassified as \texttt{\small TextViews}.  Such misclassifications in the GUI-hierarchy would be trivial for experienced Android developers to fix in the generated app while the GUI-hierarchy and boilerplate code would be automatically generated by ReDraw. The strong performance of the CNN-based classifier provides a solid base for the application generation procedure employed by ReDraw.  Based on these results, we answer \textbf{RQ$_1$:}
	\begin{center}
		\fbox{
			\begin{minipage}[t]{0.9\linewidth}
				\noindent \textbf{RQ$_1$: ReDraw's CNN-based GUI-component classifier was able to achieve a high average precision (91\%) and outperform the baseline BOVW approach's average precision (65\%).}
			\end{minipage}
		}
	\end{center}

\begin{figure*}[t]
	\centering
	\includegraphics[width=\textwidth]{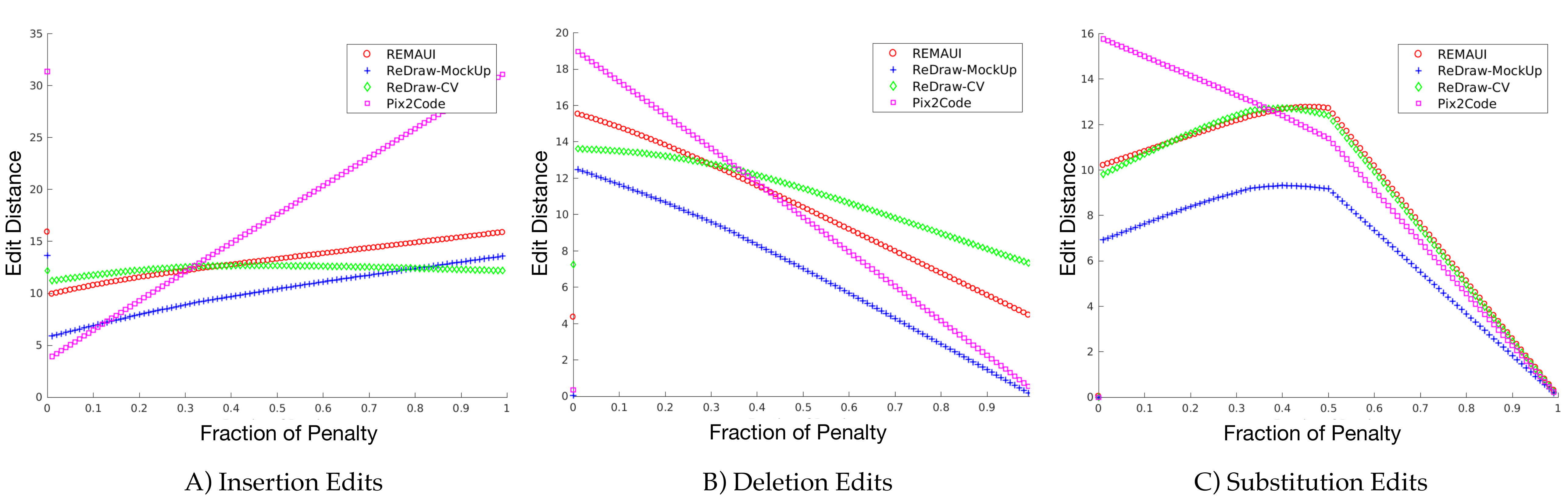}
	\vspace{-1em}
	\caption{Hierarchy similarities based on edit distances}
	\label{fig:hierarchy-results}
	\vspace{-1.5em}
\end{figure*}

\vspace{-0.3em}
\subsection{RQ$_2$ Results: Hierarchy Construction}
\label{subsec:results-rq2}

	An important part of the app generation process is the automated construction of a GUI-hierarchy to allow for the proper grouping, and thus proper displaying, of GUI-components into GUI-containers.  Our evaluation of ReDraw's GUI-hierarchy construction compares against the \Remaui and \pixcode approaches by decomposing the runtime GUI-hierarchies into trees and measuring the edit distance between the generated trees and target trees (as described in Section \ref{subsec:rq2}). By varying the penalty prescribed to each edit operation, we can gain a more comprehensive understanding of the similarity of the generated GUI-hierarchies by observing, for instance, whether certain hierarchies were more or less shallow than real applications, by examining the performance of insertion and deletion edits.

The results for our comparison based on Tree edit distance are illustrated in Fig. \ref{fig:hierarchy-results} A-C.  Each graph illustrates the results for a different edit operation and the lines delineated by differing colors and shapes represent the studied approaches (\ReDraw Mock-Up or CV-based, \Remaui, or pix2code) with the edit distance (\eg closeness to the target hierarchy) shown on the y-axis and the penalty prescribed to the edit operation on the x-axis. For each of the graphs, a lower point or line indicates that a given approach was closer to the target mock-up hierarchy. The results indicate that in general, across all three variations in edit distance penalties, \ReDraw-MockUp produces hierarchies that are closer to the target hierarchies than \Remaui and pix2code.  Of particular note is that as the cost of insertion operations rises both \ReDraw-CV and \ReDraw-MockUp outperform REMAUI.  In general \ReDraw-Mockup requires fewer than ten edit operations across the three different types of operations to exactly match the target app's GUI-hierarchy.  While \ReDraws hierarchies require a few edit operations to exactly match the target, this may be acceptable in practice, as there may be more than one variation of an acceptable hierarchy.  Nevertheless, \ReDraw-Mockup is closer than other related approaches in terms of similarity to real hierarchies. 

Another observable phenomena exhibited by this data is the tendency for \Remaui and \pixcode to generate relatively \textit{shallow} hierarchies. We see that as the penalty for insertion increases, both \ReDraw-CV and \ReDraw-Mockup outperform \Remaui and pix2code. This is because ReDraw simply does not have to perform as many insertions into the hierarchy to match the ground truth. Pix2code and \Remaui are forced to add more inner nodes to the tree because their generated hierarchies are too shallow (i.e. lacking in inner nodes). From a development prototyping point of view, it is more likely easier for a developer to remove redundant nodes than it is to create new nodes, requiring them reasoning what amounts to a new hierarchy after the automated prototyping process. These results are unsurprising for the \Remaui approach, as the authors used shallowness as a proxy for suitable hierarchy construction.  However, this evaluation illustrates that the shallow hierarchies generated by \Remaui and \pixcode do match the target hierarchies as well as those generated by \ReDraw-Mockup.  While minimal hierarchies are desirable from the point of view of rendering content on the screen, we find that REMAUI's hierarchies tend to be dramatically more shallow compared to \ReDraws which exhibit higher similarity to real hierarchies.  Another important observation is that the substitution graph illustrates the general advantage that the CNN-classifier affords during hierarchy construction.  \ReDraw-Mockup requires far fewer substitution operations to match a given target hierarchy than \Remaui, which is at least in part due to \ReDraws ability to properly classify GUI-components, compared to the text/image binary classification afforded by \Remaui.  From these results, we can answer \textbf{RQ$_2$:}  
	\begin{center}
		\fbox{
			\begin{minipage}[t]{0.9\linewidth}
				\noindent \textbf{\textbf{RQ$_2$:} \ReDraw-MockUp is capable of generating GUI-hierarchies closer in similarity to real hierarchies than \Remaui or pix2code. This signals that ReDraw's hierarchies can be utilized by developers with low effort.}
			\end{minipage}
		}
	\end{center}

\subsection{RQ$_3$ Results: Visual Similarity}
\label{subsec:results-rq3}

\begin{figure*}[t]
	\centering

	\includegraphics[width=0.8\linewidth]{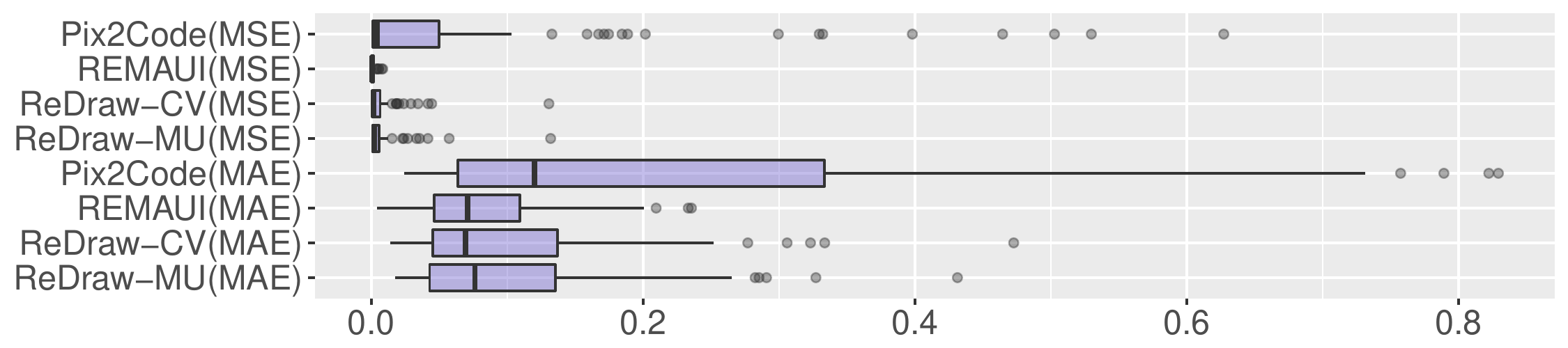}

	\caption{Pixel-based mean average error and mean squared error of screenshots: \ReDraw, \Remaui, and \pixcode apps}
	\label{fig:visual-results}
\end{figure*}

 An effective GUI-prototyping approach should be capable of generating apps that are visually similar to the target mock-up artifacts.  We measured this by calculating the MAE and MSE across all pixels in screenshots from generated apps for ReDraw-MockUp, \ReDraw-CV, \Remaui, and \pixcode (Fig. \ref{fig:visual-results}.) compared to the original app screenshots.  This figure depicts a box-and-whisker plot with points corresponding to a measurement for each of the studied 83 subject applications.  The black bars indicate mean values.  In general, the results indicate that all approaches generated apps that exhibited high overall pixel-based similarity to the target screenshots.  \ReDraw-CV outperformed both REMAUI and \pixcode in MAE, \revision{whereas all approaches exhibited very low MSE, with REMAUI very slightly outperforming both ReDraw variants.} The apps generated by \pixcode exhibit a rather large variation from the target screenshots used as input.  This is mainly due to the artificial nature of the training set utilized by \pixcode which in turn generates apps only with a relatively rigid, pre-defined set of components.  The results of the Mann-Whitney test reported in Table \ref{tab:pixel-based-mae} \& \ref{tab:pixel-based-mse} illustrate wether the similarity between each combination of approaches was statistically significant.  For MAE, we see that when \ReDraw-CV and \ReDraw-Mockup are compared to \Remaui, the results are not statistically significant, however, when examining the MSE for these same approaches the result is statistically significant with a medium effect effect size according to the Cliff's delta measurement.  \revision{Thus, it is clear that on average \ReDraw and \Remaui both generate prototype applications that are closely similar to a target visually, with \Remaui outperforming \ReDraw in terms of MSE to a statistically significant degree (with the overall MSE being extremely low $<0.007$ for both approaches) and \ReDraw outperforming \Remaui in terms of average MAE (although not to a statistically significant degree).}  \revision{This is encouraging, given that \Remaui directly copies images of components (including those that are not images, like buttons) and generates text-fields. Reusing images for all non-text components is likely to lead to more visually similar (but less functionally accurate) apps than classifying the proper component type and inferring the stylistic details of such components.}  When comparing both variants of \ReDraw and \Remaui to pix2code, the results are all statistically significant, with ranging effect sizes.  Thus, both \ReDraw and \Remaui outperform pix2code in terms of generating prototypes that are visually similar to a target.

\begin{table}[tb]
	\vspace{-0.3cm}
	\caption{Pixel-based comparison by MAE: Mann-Whitney test ($p$-value) and Cliff's Delta ($d$).}
	\label{tab:pixel-based-mae}
	\centering
	\resizebox{\linewidth}{!}{
		\begin{tabular}{lrr}\hline
			Test & $p$-value & $d$ \\ \hline
			\textbf{ReDrawâMU} \emph{vs} \textbf{ReDrawâCV} & 0.835  & 0.02 (Small)\\
			\textbf{ReDrawâMU} \emph{vs} \textbf{REMAUI} & 0.542  & 0.06 (Small)\\
			\textbf{ReDrawâMU} \emph{vs} \textbf{pix2Code} & $<$ 0.0002  & -0.34 (Medium)\\
			\textbf{pix2Code} \emph{vs} \textbf{ReDrawâCV} & $<$ 0.0001  & 0.35 (Medium)\\
			\textbf{pix2Code} \emph{vs} \textbf{REMAUI} & $<$ 0.0001  &  0.39 (Medium)\\
			\textbf{REMAUI} \emph{vs} \textbf{ReDrawâCV} & 0.687  & -0.04 (Small)\\
			\hline
		\end{tabular}
	}
\end{table}

	While in general the visual similarity for apps generated by \ReDraw is high, there are instances where \Remaui outperformed our approach.  Typically this is due to instances where \ReDraw misclassifies a small number of components that cause visual differences. For example, a button may be classified and rendered as a switch in rare cases.  However, \Remaui does not suffer from this issue as all components deemed not to be text are copied to the generated app as an image.  While this occasionally leads to more visually similar apps, the utility is dubious at best, as developers will be required to add proper component types, making extensive edits to the GUI-code.  Another instance that caused some visual inconsistencies for \ReDraw was text overlaid on top of images.  In many cases, a developer might overlay a snippet of text over an image to create a striking effect (\eg Netflix often overlays text across movie-related images).  However, this can cause an issue for \ReDraws prototyping methodology.  During the detection process, \ReDraw recognizes images and overlaid text in a mockup.  However, given the constraints of our evaluation, \ReDraw simply re-uses the images contained within screenshot as is, which might include overlaid text.  Then, ReDraw would render a \texttt{\small TextView} or \texttt{\small EditText} over the image \textit{which already includes the overlaid text} causing duplicate lines of text to be displayed.  In a real-world prototyping scenario, such issues can be mitigated by designers providing ``clean" versions of the images used in a mockup, so that they could be utilized in place of ``runtime" images that may have overlaid text.  Overall, the performance of \ReDraw is quite promising in terms of the visual fidelity of the prototype apps generated, with the potential for improvement if adopted into real design workflows.  
 
\begin{figure*}
	\centering
	\includegraphics[width=0.80\paperwidth]{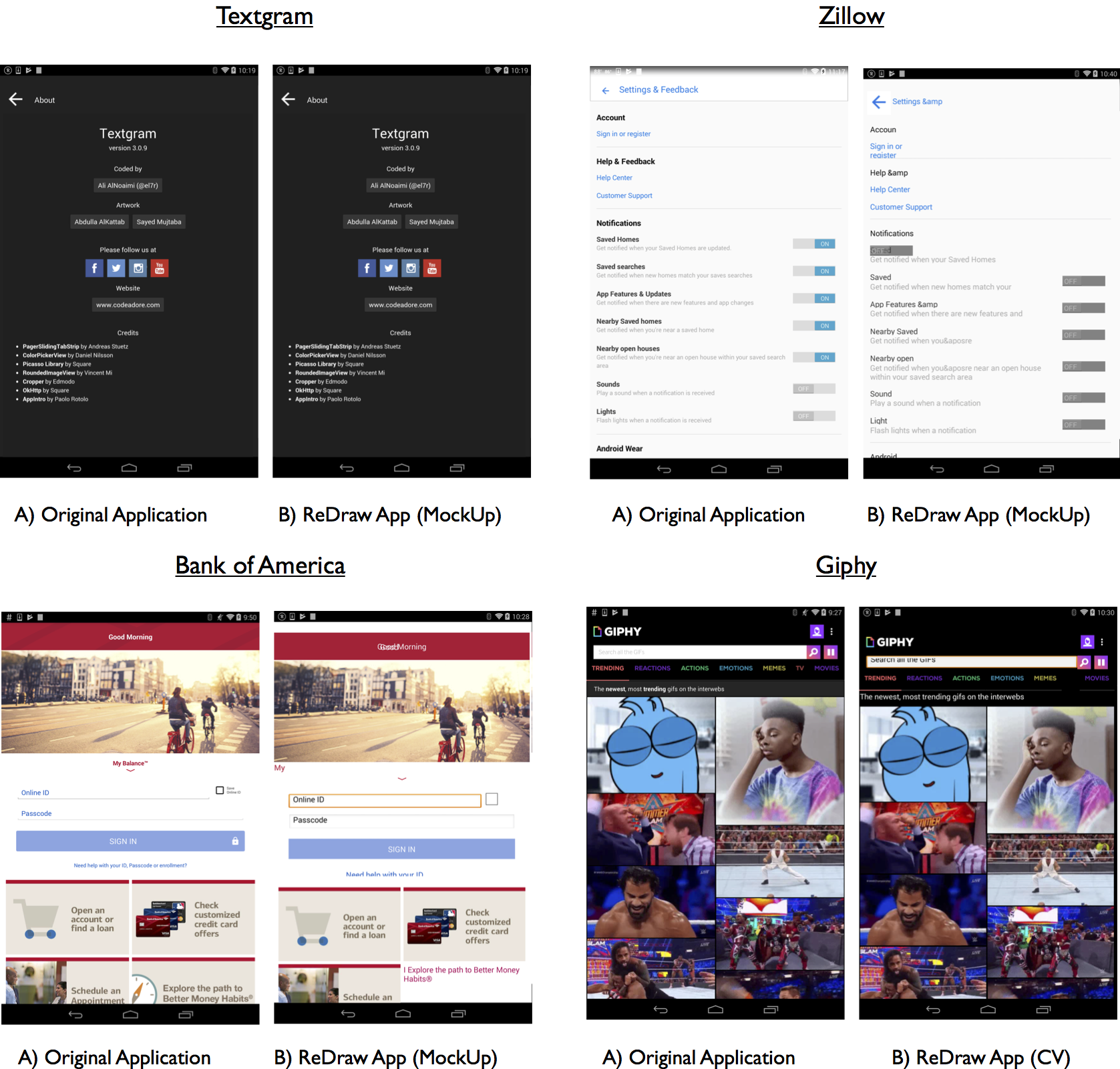}
	\vspace{-0.2em}
	\caption{Examples of apps generated with \ReDraw exhibiting high visual and structural similarity to target apps}
	\label{fig:app-examples}
	\vspace{-1.5em}
\end{figure*}

 We illustrate some of the more successful generated apps (in terms of visual similarity to a target screenshot) in Fig. \ref{fig:app-examples}; screenshots and hierarchies for all generated apps are available in a dataset in our online appendix \cite{appendix}.  In summary, we can answer \textbf{RQ$_3$} as follows: 
\begin{center}
	\fbox{
		\begin{minipage}[t]{0.9\linewidth}
			\noindent \textbf{RQ$_3$: The apps generated by ReDraw exhibit high visual similarity compared to target screenshots.}
		\end{minipage}
	}
\end{center}

\begin{table}[tb]
\vspace{-0.3cm}
	\caption{Pixel-based comparison by MSE: Mann-Whitney test ($p$-value) and Cliff's Delta ($d$).}
	\label{tab:pixel-based-mse}
	\centering
	\resizebox{\linewidth}{!}{
		\begin{tabular}{lrr}\hline
			Test & $p$-value & $d$ \\ \hline
			\textbf{ReDrawâMU} \emph{vs} \textbf{ReDrawâCV} & 0.771  & 0.03 (Small)\\
			\textbf{ReDrawâMU} \emph{vs} \textbf{REMAUI} & $<$ 0.0001  & 0.45 (Medium)\\
			\textbf{ReDrawâMU} \emph{vs} \textbf{pix2Code} & $<$ 0.003  & -0.27 (Small)\\
			\textbf{pix2Code} \emph{vs} \textbf{ReDrawâCV} & $<$ 0.002  & 0.28 (Small)\\
			\textbf{pix2Code} \emph{vs} \textbf{REMAUI} & $<$ 0.0001  &  0.61 (Large)\\
			\textbf{REMAUI} \emph{vs} \textbf{ReDrawâCV} & $<$0.0001  & -0.42 (Medium)\\
			\hline
		\end{tabular}
	}
\end{table}

\vspace{-1.5em}
\subsection{RQ$_4$ Results: Industrial Applicability}
\label{subsec:results-rq4}

	To understand the applicability of \ReDraw from an industrial prospective we conducted a set of semi-structured interviews with a front-end Android developer @Google, a mobile designer @Huawei, and a mobile researcher @Facebook. We asked them four questions (see Sec. \ref{sec:study}) related to (i) the applicability of \ReDraw, (ii) aspects of \ReDraw they found beneficial, and (iii) areas for improvement.

\vspace{-0.5em}
	\subsubsection{Front End Android Developer @Google}  The first individual works mostly on Google's search products, and his team practices the process of mock-up driven development, where developers work in tandem with a dedicated UI/UX team. Overall, the developer was quite positive about \ReDraw explaining that it could help to improve the process of writing a new Android app activity from scratch, however, he noted that \textit{``It's a good starting point... From a development standpoint, the thing I would appreciate most is getting a lot of the boilerplate code done [automatically]".} In the ``boilerplate" code statement, the developer was referring to the large amount of layout and style code that must be written when creating a new activity or view.  He also admitted that this code is typically written by hand stating, \textit{``I write all my GUI-code in xml, I don't use the Android Studio editor, very few people use it"}. He also explained that this GUI-code is time-consuming to write and debug stating, \textit{``If you are trying to create a new activity with all its components, this can take hours"}, in addition to the time required for the UI/UX team to verify proper implementation.  The developer did state that some GUI-hierarchies he examined tended to have redundant containers, but that these can be easily fixed stating, \textit{``There are going to be edge cases for different layouts, but these are easily fixed after the fact"}. 

	The aspect of \ReDraw that this developer saw the greatest potential for, is its use in an evolutionary context.  During the development cycle at Google, the UI/UX team will often propose changes to existing apps, whose GUI-code must be updated accordingly.  The developer stated that \ReDraw had the potential to aid this process: \textit{``The key thing is fast iteration. A developer could generate the initial view [using ReDraw], clean up the layouts, and have a working app. If a designer could upload a screenshot, and without any other intervention [ReDraw] could update the [existing] xml this would be ideal."} The developer thought that if \ReDraw was able to detect existing GUI-components in a prior app version, and update the layouts and styles of these components according to a screenshot, generating new components as necessary, this could greatly improve the turn around time of GUI-changes and potentially increase quality.  He even expressed optimism that the approach could learn from developer corrections on generated code over time, stating \textit{``It would be great if you could give it [ReDraw] developer fixes to the automatically generated xml and it could learn from this."}

\vspace{-0.5em}
\subsubsection{Mobile UI/UX Designer @Huawei} 
\vspace{-0.3em}
We also interviewed a dedicated UI/UX designer at Huawei, with limited programming experience.  His primary job is to create mock-up artifacts that stipulate designs of mobile apps, communicate these to developers, and ensure they are implemented to spec.  This interview was translated from Chinese into English.  This designer also expressed interest in \ReDraw, stating that the visual similarity of the apps was impressive for an automated approach, \textit{``Regarding visual, I feel that it's very similar"}, and that such a solution would be sought after at Huawei, \textit{``If it [a target app] can be automatically implemented after the design, it should be the best design tool [we have]"}. While this designer does not have extensive development experience, he works closely with developers and stated that the quality of the reusability of the code is a key point for adoption, \textit{``In my opinion, for the developers it would be ideal if the output code can be reused"}. This is promising as \ReDraw was shown to generate GUI-hierarchies that are comparatively more similar to real apps than other approaches.

\vspace{-1em}
\subsubsection{Mobile Researcher @Facebook} The last participant was a mobile systems researcher at Facebook.  This participant admitted that Facebook would most likely not use \ReDraw in its current state, as they are heavily invested in the React Native ecosystem. However, he saw the potential of the approach if it were adopted for this domain, stating \textit{``I can see this as a possible tool to prototype designs"}.  He was impressed by the visual similarity of the apps, stating, \textit{``The visual similarity seems impressive"}. 

In the end, we can answer \textbf{RQ$_4$:}  
\begin{center}
	\fbox{
		\begin{minipage}[t]{0.9\linewidth}
			\noindent \textbf{RQ$_4$: \ReDraw has promise for application into industrial design and development workflows, particularly in an evolutionary context.  However, modifications would most likely have to be made to fit specific workflows and prototyping toolchains.}
		\end{minipage}
	}
\end{center}

\section{Limitations \& Threats to Validity}
\label{sec:limitations-threats}

In this section we describe some limitations and possible routes for future research in automated software prototyping, along with potential threats to validity of our approach and study.

\subsection{Limitations and Avenues for Future Work} While \ReDraw is a powerful approach for prototyping GUIs of mobile apps, it is tied to certain practical limitations, some of which represent promising avenues for future work in automated software prototyping.  First, \ReDraw is currently capable of prototyping a single screen for an application, thus if multiple screens for a single app are desired, they must be prototyped individually and then manually combined into a single application.  It would be relatively trivial to modify the approach and allow for multiple screens within a single application with a simple swipe gesture to switch between them for software demo purposes however, we leave this a future work.  Additionally, future work might examine a learning-based approach for prototyping and linking together multiple screens, learning common app transitions via dynamic analysis and applying the learned patterns during prototyping.   

	Second, the current implementation of KNN-hierarchy construction is tied to the specific screen size of the devices used during the data-mining and automated dynamic analysis.  However, it is possible to utilize display independent pixel (\texttt{\small{dp}}) vslues to generalize this algorithm to function independently of screen size, we leave this as future work.  

	\revision{Third, as discussed in Section \ref{subsec:inferring-styles}, \ReDraw is currently limited to detecting and assembling a distinct set of stylistic details from mock-up artifacts including: (i) background colors; (ii) font colors, and (iii) font sizes.  \ReDraw was able to produce prototype applications that exhibited high visual similarity to target screenshots using only these inferences. However, a promising area for future work on automated prototyping of software GUIs involves expanding the stylistic details that can be inferred from a target mock-up artifact. Future work could perform more detailed studies on the visual properties of individual components from prototype screens generated from screenshots of open source apps. This study could then measure how well additional inferred styles of individual components match the original developer implemented components.} 

	Our current CNN classifier is capable of classifying incoming images into one of 15 of the most popular Android GUI-components. Thus, we do not currently support certain, rarely used component types.  Future work could investigate network architectures with more capacity (\eg deeper architectures) to classify larger numbers of component types, or even investigate emerging architectures such as Hierarchical CNNs \cite{Wang:15}. Currently, \ReDraw requires two steps for \textit{detecting} and \textit{classifying} components, however, future approaches could examine the applicability of CNN-based object detection networks \cite{Ren:NIPS15,Girshick:CVPR14} that may be capable of performing these two steps in tandem. 

\subsection{Internal Validity} Threats to \textit{internal validity} correspond to unexpected factors in the experiments that may contribute to observed results.  One such threat stems from our semi-structured interview with industrial developers.  While evaluating industrial applicability of \ReDraw, threats may arise from our manual reverse engineering of Sketch mock-ups.  However, we applied a state of art tool for detecting design violations in GUIs \cite{Moran:ICSE18} in order to ensure their validity, sufficiently mitigating this threat.  

\revision{During our experimental investigation of RQ$_2$-RQ$_4$, we utilized two different types of mock-up artifacts, (i) images of existing application screens (RQ$_2$ \& RQ$_3$, and (ii) reverse engineered mock-ups from existing application screens. The utilization of these artifacts represents a threat to internal validity as they are used as a proxy for real mock-up artifacts. However, real mock-ups created during the software design process may exhibit some unique characteristics not captured by these experimental proxies. For example, software GUI designs can be highly fluid, and oftentimes, may not be complete when handed off to a developer for implementation. Furthermore, real mock-ups may stipulate a design that cannot be properly instantiated in code (\ie unavailable font types, components organized in spatial layouts that are not supported in code). We acknowledge that our experiments do not measure the performance of \ReDraw in such cases. However, collecting real mock-up artifacts was not possible in the scope of our evaluation, as they are typically not included in the software repositories of open source applications. We performed a search for such artifacts on all Android projects hosted on GitHub as of Spring 2017, and found that no repository contained mock-ups created using Sketch.}  As stated earlier, it was not practically feasible to reverse-engineer mock-ups for all 83 applications utilized in our dataset for these experiments. Furthermore, these screenshots represent production-grade app designs that are used daily by millions of users, thus we assert that these screenshots and mock-ups represent a reasonable evaluation set for \ReDraw. We also did not observe any confounding results when applying \ReDraw to our nine reverse engineered Sketch mock-ups, thus we assert that this threat to validity is reasonably mitigated.

	Another potential confounding factor is our dataset of labeled components used to train, validate, and test the CNN.  To help ensure a correct, coherent dataset, we applied several different data filtering, cleaning, and augmentation techniques, inspired by past work on image classification using CNNs described in detail in Sec. \ref{subsec:impl-classification}.  Furthermore, we utilized the \texttt{\small uiautomator} tool included in the Android SDK, which is responsible for reporting information about runtime GUI-objects, and is generally accurate as it is tied directly to Android sub-systems responsible for rendering the GUI.  To further ensure the validity of our dataset, we randomly sampled a statistically significant portion of our dataset and manually inspected the labeled images \textit{after} our data-cleaning process was applied. We observed no irregularities and thus mitigating a threat related to the quality of the dataset.  \revision{It is possible that certain components can be specifically styled by developers to look like other components (\eg a textview styled to look like a button) that could impact the CNN component classifications.  However, our experiments illustrate that in our real-world dataset overall accuracy is still high, suggesting that such instances are rare.}  Our full dataset and code for training the CNN are available on \ReDraws website to promote reproducibility and transparency \cite{appendix}.

	\revision{During our evaluation of \ReDraws ability to generate suitable GUI-hierarchies, we compared them against the actual hierarchies of the original target applications. However, it should be noted, that the notion of a correct hierarchy may vary between developers, as currently, there is no work that empirically quantifies what constitutes a \textit{good} GUI-hierarchy for Android applications. For instance, some developers may prefer a more rigid layout with fewer container components, whereas others may prefer more components to ensure that their layout is highly reactive across devices.  We compared the hierarchies generated by ReDraw to the original apps to provide an objective measurement on actual implementations of popular apps, which we assert provides a reasonable measurement of the effectiveness of \ReDraws hierarchy construction algorithm. It should also be noted that performing this comparison on apps of different popularity levels may yield different results. We chose to randomly sample the apps from the top-10 of each Google Play category, to investigate wether \ReDraw is capable of assembling GUI-hierarchies of ``high-quality'' apps as measured by popularity.}

\subsection{Construct Validity} Threats to \textit{construct validity} concern the operationalization of experimental artifacts.  One potential threat to construct validity lies in our reimplementation of the \Remaui tool.  As stated earlier, the original version of REMAUI's web tool was not working at the time of writing this paper. We reimplemented REMAUI according to the original description in the paper, however we excluded the list generation feature, as we could not reliably re-create this feature based on the provided description. While our version may vary slightly from the original, it still represents an unsupervised CV-based technique against which we can compare \ReDraw. Furthermore, we offer our reimplementation of \Remaui (a Java program with opencv \cite{opencv} bindings) as an open source project \cite{appendix} to facilitate reproducibility and transparency in our experimentation. 

	Another potential threat to construct validity lies in our operationalization of the pix2code project.  We closely followed the instructions given in the \texttt{\small README} of the pix2code project on GitHub to train the machine translation model and generate prototype applications.  Unfortunately, the dataset used to train this model differs from the large scale dataset used to train the \ReDraw CNN and inform the KNN-hierarchy construction, however, this is due to the fact pix2code requires the source code of training applications and employs a custom domain specific language, leading to incompatibilities to our dataset.  We include the pix2code approach as a comparative baseline in this paper as it is one of the few approaches aimed at utilizing ML to perform automated GUI prototyping, and utilizes an architecture based purely upon neural machine translation, differing from our architecture.  However, it should be noted that if trained on a proper dataset, with more advanced application assembly techniques, future work on applying machine translation to automated GUI-prototyping may present better results than those reported in this paper for pix2code.

\subsection{External Validity}  Threats to \textit{external validity} concern the generalization of the results. \revision{While we implemented \ReDraw for Android and did not measure its generalization to other domains, we believe the general architecture that we introduce in this paper could transfer to other platforms or types of applications.}  This is tied to the fact that other GUI-frameworks are typically comprised sets of varying types of widgets, and GUI-related information can be automatically extracted via dynamic analysis using one of a variety of techniques including accessibility services \cite{Grechanik:ICSTW09}. \revision{While there are likely challenges that will arise in other domains, such as a higher number of component types and the potential for an imbalanced dataset, we encourage future work on extending ReDraw to additional domains.}

\revision{\ReDraw relies upon automated dynamic analysis and scraping of GUI-metadata from explored application screens to gather training data for its CNN-based classifier. However, it is possible that other application domains do not adequately expose such metadata in an easily accessible manner. Thus, additional engineering work or modification of platforms may be required in order to effectively extract such information. If information for a particular platform is difficult to extract, future work could look toward transfer learning as a potential solution. In other words, the weights for a network trained on GUI metadata that is easily accessible (\eg from Android apps) could then be fine-tuned on a smaller number of examples from another application domain, potentially providing effective results.}


\section{Conclusion \& Future Work}
\label{sec:conclusion}

In this paper we have presented a data-driven approach for automatically prototyping software GUIs, and an implementation of this approach in a tool called \ReDraw for Android.  A comprehensive evaluation of \ReDraw demonstrates that it is capable of (i) accurately detecting and classifying GUI-components in a mock-up artifact, (ii) generating hierarchies that are similar to those that a developer would create, (iii) generating apps that are visually similar to mock-up artifacts, and (iv) positively impacting industrial workflows.  In the future, we are planning on exploring CNN architectures aimed at object detection to better support the detection task. Additionally, we are planning on working with industrial partners to integrate \ReDraw, and our broader prototyping approach, into their workflows.

\section*{Acknowledgment}

We would like to thank Ben Powell, Jacob Harless, Ndukwe Iko, and Wesely Hatin from William \& Mary for their assistance on the component of our approach that generates GUI code. We would also like to thank Steven Walker and William Hollingsworth for their assistance in re-implementing the REMAUI approach. Finally, we would like to thank Martin White and Nathan Owen for their invaluable guidance at the outset of this project and the anonymous reviewers for their insightful comments which greatly improved this paper. This work is supported in part by the NSF CCF-1525902 grant. Any opinions, findings, and conclusions expressed herein are the authors’ and do not necessarily reflect those of the sponsors.




%

\bibliography{ms}
\bibliographystyle{IEEEtran}

%
\vspace{-3em}

\begin{IEEEbiography}[{\includegraphics[width=1in,height=1.25in,clip,keepaspectratio]{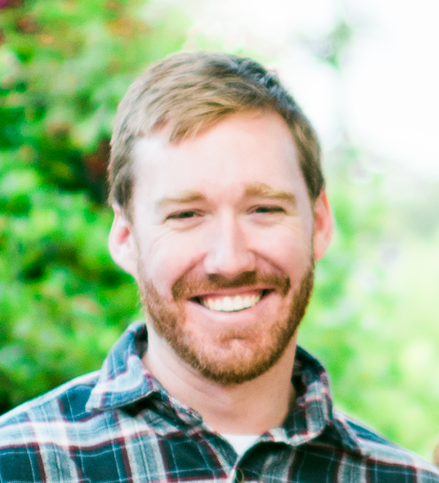}}]{Kevin Moran} is currently a Post-Doctoral researcher in the Computer Science Department at the College of William \& Mary. He is also a member of the SEMERU research group. He graduated with a B.A. in Physics from the College of the Holy Cross in 2013 and an M.S. degree from William \& Mary in August of 2015.  He received a Ph.D. degree from William \& Mary in August 2018. His main research interest involves facilitating the processes of software engineering, maintenance, and evolution with a focus on mobile platforms. He has published in several top peer-reviewed software engineering venues including: ICSE, ESEC/FSE, TSE, USENIX, ICST, ICSME, and MSR. He was recognized as the second-overall graduate winner in the ACM Student Research competition at ESEC/FSE15. Moran is a member of IEEE and ACM and has served as an external reviewer for ICSE, FSE, ASE, ICSME, APSEC, and SCAM. More information available at \url{http://www.kpmoran.com}.
\end{IEEEbiography}

\vspace{-4em}

\begin{IEEEbiography}[{\includegraphics[width=1in,height=1.25in,clip,keepaspectratio]{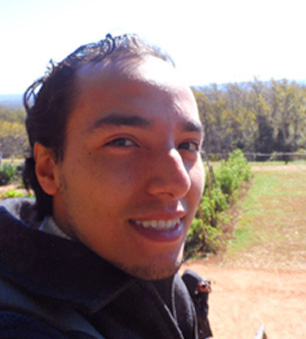}}]{Carlos Bernal-C\'ardenas} received the BS degree in systems engineering from the Universidad Nacional de Colombia in 2012 and his M.E. in Systems and Computing Engineering in 2015. He is currently Ph.D. candidate in Computer Science at the College of William \& Mary as a member of the SEMERU research group advised by Dr Denys Poshyvanyk. His research interests include software engineering, software evolution and maintenance, information retrieval, software reuse, mining software repositories, mobile applications development, and user experience. He has published in several top peer-reviewed software engineering venues including: ICSE, ESEC/FSE, ICST, and MSR.  He has also received the ACM SigSoft Distinguished paper award at ESEC/FSE'15.  Bernal-C\'ardenas is a student member of IEEE and ACM and has served as an external reviewer for ICSE, ICSME, FSE, APSEC, and SCAM. More information is available at \url{http://www.cs.wm.edu/~cebernal/}.
\end{IEEEbiography}

\vspace{-4em}

\begin{IEEEbiography}[{\includegraphics[width=1in,height=1.25in,clip,keepaspectratio]{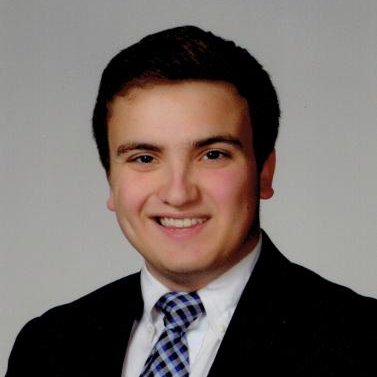}}]{Michael Curcio} is an undergraduate student in the Computer Science Department at the College of William \& Mary.  He is currently a member of the SEMERU research group and is pursuing an undergraduate honors thesis on the topic of automating software design workflows. His research interests lie in applications of deep learning to software engineering and design tasks. Curcio is an IEEE student member.
\end{IEEEbiography}

\vspace{-4em}

\begin{IEEEbiography}[{\includegraphics[width=1in,height=1.25in,clip,keepaspectratio]{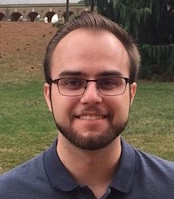}}]{Richard Bonett} is a MS/PhD student at The College of William \& Mary and a member of the SEMERU research group. He graduated from The College of William \& Mary with a B.S. in Computer Science in Spring 2017. His primary research interests lie in Software Engineering, particularly in the development and evolution of mobile applications. Bonett has recently published at MobileSoft'17. More information is available at \url{http://www.cs.wm.edu/~rfbonett/}.
\end{IEEEbiography}

\vspace{-4em}

\begin{IEEEbiography}[{\includegraphics[width=1in,height=1.25in,clip,keepaspectratio]{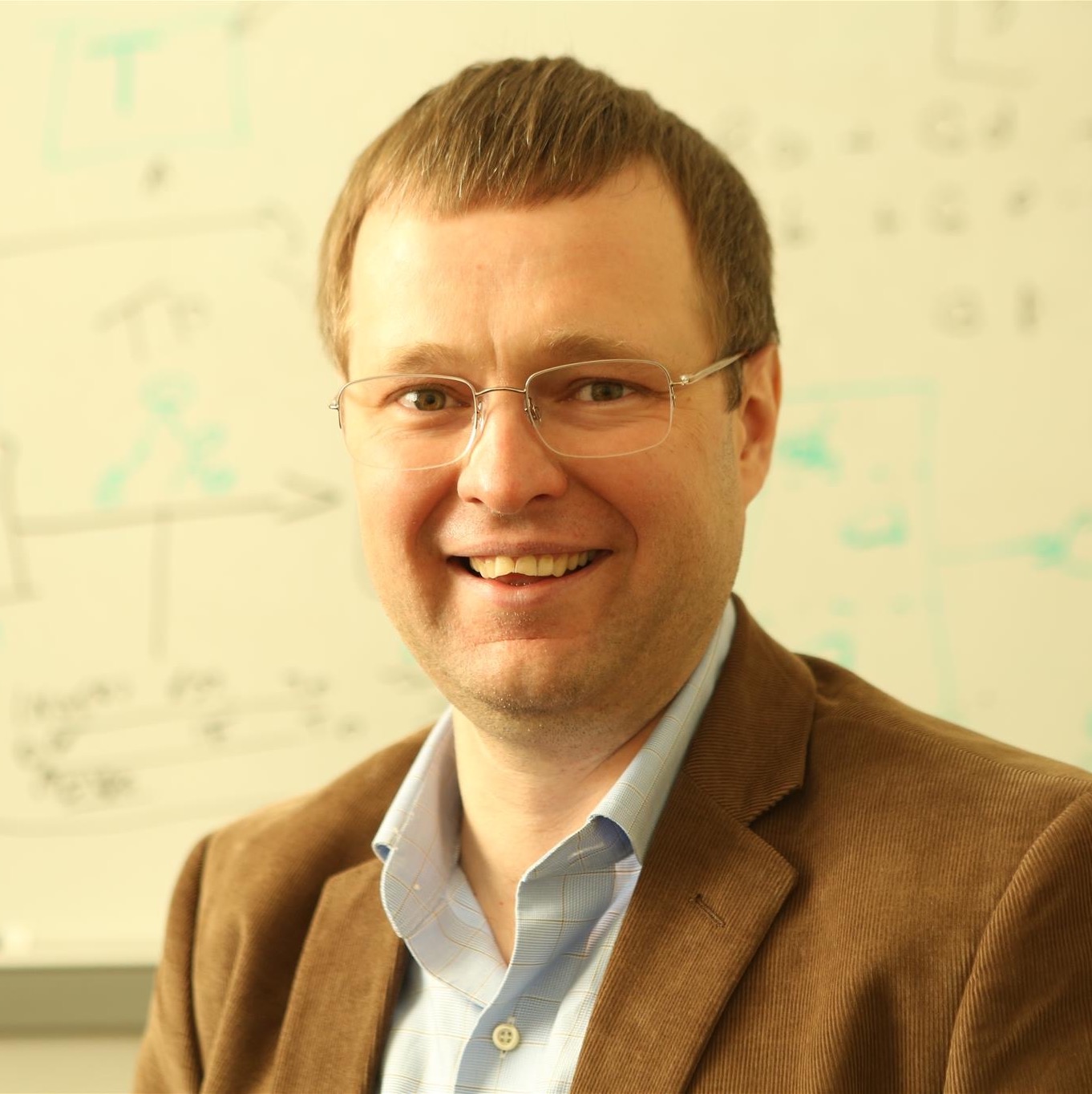}}]{Denys
Poshyvanyk} is the Class of 1953 Term Distinguished Associate Professor of Computer Science at the College of William \& Mary in Virginia. He received the MS and MA degrees in Computer Science from the National University of Kyiv-Mohyla Academy, Ukraine, and Wayne State University in 2003 and 2006, respectively. He received the PhD degree in Computer Science from Wayne State University in 2008. He served as a program co-chair for ICSME'16, ICPC'13, WCRE'12 and WCRE'11. He currently serves on the editorial board of IEEE Transactions on Software Engineering (TSE), Empirical Software Engineering Journal (EMSE, Springer) and Journal of Software: Evolution and Process (JSEP, Wiley). His research interests include software engineering, software maintenance and evolution, program comprehension, reverse engineering, software repository mining, source code analysis and metrics. His research papers received several Best Paper Awards at ICPC'06, ICPC'07, ICSM'10, SCAM'10, ICSM'13 and ACM SIGSOFT Distinguished Paper Awards at ASE'13, ICSE'15, ESEC/FSE'15, ICPC'16 and ASE'17. He also received the Most Influential Paper Awards at ICSME'16 and ICPC'17. He is a recipient of the NSF CAREER award (2013).  He is a member of the IEEE and ACM. More information available at: http://www.cs.wm.edu/~denys/.
\end{IEEEbiography}

\balance




\end{document}